\documentclass[%
reprint,aps,prc,
superscriptaddress,
showpacs,preprintnumbers,
amsmath,
amssymb,
twocolumn,
floatfix,  
]{revtex4-2}
\usepackage{array}[=2016-10-06]

\usepackage{lineno}  
\usepackage{lipsum}  

\usepackage{graphicx} 
\usepackage{dcolumn}
\usepackage{bm}
\usepackage{lmodern} 
\usepackage{amsfonts}
\usepackage{amssymb}
\usepackage{gensymb}
\usepackage{caption}
\usepackage{sidecap}
\usepackage{hyperref}
\usepackage{xcolor}
\usepackage{float}
\usepackage{multirow}
\usepackage{caption}
\usepackage{subcaption}
\usepackage{tabularray}  
\usepackage{array}       
\usepackage{enumitem} 
\usepackage[normalem]{ulem} 
\usepackage{siunitx}
\usepackage{comment}

\usepackage[font=footnotesize]{caption}

\newcommand{\softsh}[1]{\texttt{#1}}
\newcommand{\soft}[1]{\texttt{#1}\,}

\newcommand{\lisepp}{\soft{LISE$^{++}$}}
\newcommand{\liseppcute}{\soft{LISE$^{++}_{cute}$}}
\newcommand{\liseppsh}{\softsh{LISE$^{++}$}}

\newcommand{\nuc}[2]{\hbox{$^{#1}$#2}}

\begin{document}

\preprint{APS. Version 2.3}

\title{Observation of \texorpdfstring{$^{71}$Cr}{71Cr} and expanded cross-section measurements and analysis of \texorpdfstring{$^{82}$Se}{82Se} fragmentation}


\newcommand{\aMSUphys}{\affiliation{Department of Physics and Astronomy, Michigan State University, East Lansing, MI 48824, USA}}
\newcommand{\aMSUchem}{\affiliation{Department of Chemistry, Michigan State University, East Lansing, MI 48824, USA}}
\newcommand{\aFRIB}{\affiliation{Facility for Rare Isotope Beams, Michigan State University, East Lansing, MI 48824, USA}}

\newcommand{\aLLNL}{\affiliation{Lawrence Livermore National Laboratory, Livermore, CA 94550, USA}}
\newcommand{\aUTK}{\affiliation{Department of Physics and Astronomy, University of Tennessee, Knoxville, TN 37996, USA}}
\newcommand{\aZagreb}{\affiliation{Department of Physics, Faculty of Science, University of Zagreb, Bijeni\v{c}ka c. 32, 10000 Zagreb, Croatia}}
\newcommand{\aORNL}{\affiliation{Physics Division, Oak Ridge National Laboratory, Oak Ridge, TN 37831, USA}}
\newcommand{\aFSU}{\affiliation{Department of Physics, Florida State University, Tallahassee, FL 32306, USA}}
\newcommand{\aCMU}{\affiliation{Department of Physics, Central Michigan University, Mt Pleasant, MI 48858, USA}}
\newcommand{\aUNC}{\affiliation{Department of Physics and Astronomy, University of North Carolina, Chapel Hill, NC 27599, USA}}
\newcommand{\aTri}{\affiliation{Triangle Universities Nuclear Laboratory, Duke University, Durham, NC 27708, USA}}
\newcommand{\aANL}{\affiliation{Physics Division, Argonne National Laboratory, Argonne, IL 60439, USA}}
\newcommand{\aLBNL}{\affiliation{Nuclear Science Division, Lawrence Berkeley National Laboratory, Berkeley, CA 94720, USA}}
\newcommand{\aMissU}{\affiliation{Department of Physics and Astronomy, Mississippi State University, Mississippi State, MS 39762, USA}}
\newcommand{\aPNNL}{\affiliation{Pacific Northwest National Laboratory, Richland, WA 99354, USA}}
\newcommand{\aUWar}{\affiliation{Faculty of Physics, University of Warsaw, PL-02-093 Warsaw, Poland}}
\newcommand{\aUTokyo}{\affiliation{Center for Nuclear Study, University of Tokyo, Wako, Saitama 351-0198, Japan}}
\newcommand{\aTTU}{\affiliation{Physics Department, Tennessee Technological University, Cookeville, TN 38505, USA}}
\newcommand{\aOU}{\affiliation{Department of Physics and Astronomy, Ohio University, Athens, OH 45701, USA}}
\newcommand{\aUCB}{\affiliation{University of California, Berkeley, CA 94704, USA}}
\newcommand{\aUMD}{\affiliation{Department of Chemistry, University of Maryland, College Park, MD 20742, USA}}
\newcommand{\aHIL}{\affiliation{Heavy Ion Laboratory, University of Warsaw, PL-02-093 Warsaw, Poland}}
\newcommand{\aUR}{\affiliation{Department of Physics, University of Regina, Regina, SK S4S 0A2, Canada}}
\newcommand{\aAFIT}{\affiliation{Air Force Institute of Technology, Dayton, Ohio 45433, USA}}

\newcommand{\aANU}{\affiliation{Department of Nuclear Physics and Accelerator Applications, Research School of Physics, Australian National University, Canberra, ACT 2601, Australia}}

\newcommand{\aLSU}{\affiliation{Department of Physics and Astronomy, Louisiana State University, Baton Rouge, LA, 70803, USA}}


\author{S.~Watters} \email{watterss@frib.msu.edu} \aFRIB \aMSUphys
\author{O.~B.~Tarasov} \aFRIB

\author{A.~C.~Dombos} \aFRIB 
\author{K.~Fukushima} \aFRIB 
\author{A.~Gade} \aFRIB \aMSUphys
\author{K.~Haak}  \aFRIB  \aMSUphys
\author{M.~Hausmann} \aFRIB
\author{D.~Kahl} \aFRIB
\author{D.~Kaloyanov} \aFRIB \aMSUphys
\author{E.~Kwan} \aFRIB
\author{H.~K.~Matthews} \aFRIB \aMSUphys
\author{P.~N.~Ostroumov} \aFRIB \aMSUphys
\author{M.~Portillo} \aFRIB
\author{I.~Richardson} \aFRIB \aMSUphys
\author{B.~M.~Sherrill} \aFRIB \aMSUphys
\author{M.~K.~Smith} \aFRIB

\author{S.~Ajayi} \altaffiliation{Present address: Department of Physics, Florida A\&M University, Tallahassee, Florida 32307, USA.} \aFSU 
\author{J.~M.~Allmond} \aORNL
\author{H.~Arora} \altaffiliation{Present address: Center for Exotic Nuclear Studies, Institute for Basic Science, Daejeon 34126, Republic of Korea.} \aCMU
\author{A.~D.~Ayangeakaa} \aUNC \aTri
\author{H.~C.~Berg} \aFRIB \aMSUphys
\author{J.~M.~Berkman} \aFRIB \aMSUchem
\author{D.~L.~Bleuel} \aLLNL
\author{K.~Bosmpotinis} \aFRIB \aMSUphys

\author{M.~P.~Carpenter} \aANL

\author{A.~Chester} \aFRIB
\author{J.~M.~Christie} \aUTK
\author{S.~Coil} \aFRIB \aMSUphys
\author{I.~Cox} \aUTK

\author{B.~P.~Crider} \aMissU
\author{J.~Davis} \aFSU \aPNNL
\author{A.~A.~Doetsch} \aFRIB \aMSUphys
\author{J.~G.~Duarte} \aLLNL
\author{A.~Estrade} \aCMU
\author{A.~Fijalkowska} \aUWar
\author{C.~Frantzis} \aFRIB \aMSUphys
\author{T.~Gaballah} \aFRIB \aMissU
\author{H.~Gadaria} \aFRIB \aMSUphys
\author{E.~C.~Good} \altaffiliation{Present address: Pacific Northwest National Laboratory, Richland, Washington 99354, USA} \aFRIB 
\author{T.~Gray} \aORNL
\author{R.~Grzywacz} \aUTK
\author{S.~Hanai} \aUTokyo
\author{J.~T.~Harke} \aLLNL
\author{C.~Harris} \aFRIB \aMSUphys
\author{A.~C.~Hartley} \aFRIB \aMSUphys
\author{K.~Hermansen} \aFRIB \aMSUphys
\author{D.~E.~M.~Hoff} \aLLNL
\author{D.~Hoskins III} \aUTK
\author{J.~Huffman} \aFRIB \aMSUphys
\author{R.~V.~F.~Janssens} \aUNC \aTri
\author{R.~Jain} \aFRIB \aMSUphys \aLLNL 
\author{M.~Karny} \aUWar
\author{T.~T.~King} \aORNL
\author{N.~Kitamura} \aUTK
\author{K.~Kolos} \aLLNL
\author{S.~N.~Liddick} \aFRIB \aMSUchem
\author{B.~Longfellow} \altaffiliation{Present address: Argonne National Laboratory, Lemont, Illinois 60439, USA.} \aLLNL 
\author{R.~S.~Lubna} \aFRIB
\author{S.~Lyons} \aPNNL
\author{M.~Madurga} \aUTK
\author{M.~Mogannam} \altaffiliation{Present address: Air Force Institute of Technology, Dayton, Ohio 45433, USA.} \aFRIB \aMSUchem 
\author{S.~Neupane} \aLLNL \aUTK
\author{T.~H.~Ogunbeku} \aLLNL 
\author{B.~Olree} \aMissU
\author{W.-J.~Ong} \aLLNL
\author{G.~Owens-Fryar} \aFRIB \aMSUphys
\author{J.~R.~Palomino} \aMissU
\author{M.~M.~Rajabali} \aTTU
\author{B.~C.~Rasco} \aORNL
\author{A.~L.~Richard} \aLLNL \aOU
\author{E.~K.~~Ronning} \altaffiliation{Present address: Istituto Nazionale di Fisica Nucleare, Sezione di Padova, Padua 35131, Italy.} \aFRIB \aMSUchem 
\author{G.~E.~Rose} \aUCB
\author{T.~Ruland} \aORNL
\author{K.~P.~Rykaczewski} \aORNL
\author{H.~Schatz} \aFRIB \aMSUphys
\author{N.~D.~Scielzo} \aLLNL
\author{D.~P.~Scriven} \aFRIB
\author{D.~Seweryniak} \aANL
\author{K.~Siegal} \aUTK
\author{A.~Spyrou} \aFRIB \aMSUphys
\author{M.~Stepaniuk} \aUWar \aLSU
\author{A.~E.~Stuchbery} \aANU
\author{A.~Sweet} \aLLNL
\author{V.~Tripathi} \aFSU
\author{A.~Tsantiri} \altaffiliation{Present address: Department of Physics, University of Regina, Regina, Saskatchewan S4S 0A2, Canada.} \aFRIB \aMSUphys 
\author{S.~Uthayakumaar} \aFRIB 
\author{W.~B.~Walters} \aUMD
\author{M.~Wolinska-Cichocka} \aHIL
\author{Z.~Y.~Xu} \aUTK
\author{R.~Yokoyama} \aUTokyo


\newcommand{\OT}[1]{\textcolor{blue}{#1}}
\newcommand{\OTs}[1]{\textcolor{red}{#1}}
\newcommand{\HKM}[1]{\textcolor{purple}{#1}}


\date{\today}

\begin{abstract}

Production cross sections for neutron-rich nuclei from the fragmentation of \nuc{82}{Se} were measured, including the previously unobserved isotope \nuc{71}{Cr}. Particle identification of the mass number ($A$), atomic number ($Z$), and ionic charge state ($q$) was performed event-by-event using measurements of energy loss, time of flight, magnetic rigidity, and total kinetic energy. These measurements, combined with previous \nuc{82}{Se} fragmentation data, were analyzed using an abrasion-ablation formalism and the recently developed $\Delta BE$ systematics. 
Model dependence in both approaches was quantified through a Bayesian-inspired model averaging procedure based on eleven nuclear mass models. The $\Delta BE$ systematics provide a simple means of incorporating nuclear structure information into production-rate predictions, while the abrasion-ablation analysis provides insight into the reaction mechanisms governing the production of neutron-rich nuclei. The production cross section predictions of these two approaches are in agreement for nuclei predicted to be well-bound by the mass models. Discrepancies between the two frameworks are large for nuclei near $N=48$, corresponding to pure-$Z$ removal. The agreement of $\Delta BE_{\mathrm{sys}}$ with the experimental data and the under-prediction of the AA calculations point to a potentially incomplete description of the reaction mechanism governing the production of these isotopes. The first observation of \nuc{71}{Cr} provides an out-of-sample test of the predictive capability of the $\Delta BE$ systematics: a cross section of $1.94\times 10^{-13}$ mb, predicted using the HFB-22 mass table and $\Delta BE_{\mathrm{sys}}$ coefficients determined before the observation of \nuc{71}{Cr}, is consistent with the measured value reported here: $2.20^{+2.10}_{-0.75}\times10^{-13}$ mb.

\end{abstract}

\maketitle



\section{Introduction}

The chart of nuclides contains all experimentally known combinations of protons and neutrons that form nuclei or resonance states that live long enough to be observed and studied in the laboratory. The nuclear chart, sometimes referred to as the nuclear landscape, has boundaries defined by the proton and neutron driplines. These driplines define the limits of stability against nucleon emission and are conventionally associated with negative nucleon separation energies: when the energy required to remove a neutron (proton) becomes negative, the ensemble becomes unbound to neutron (proton) emission. For heavy and superheavy nuclei, however, spontaneous fission imposes an additional constraint on nuclear existence, governed by the fission barrier and the resulting nuclear lifetime. Thus, the limits of observable nuclei depend not only on particle separation energies but also on the lifetimes associated with the available decay modes. In light of this, a different nuclear existence demarcation line can be defined based on the lifetime of the nucleus, regardless of its nucleon separation energy \cite{Thoennessen2004, EphemeralNuclei}.

\begin{figure*} [htbp!]
    \centering
    \begin{minipage}{0.75\textwidth} 
        \includegraphics[width=\textwidth]{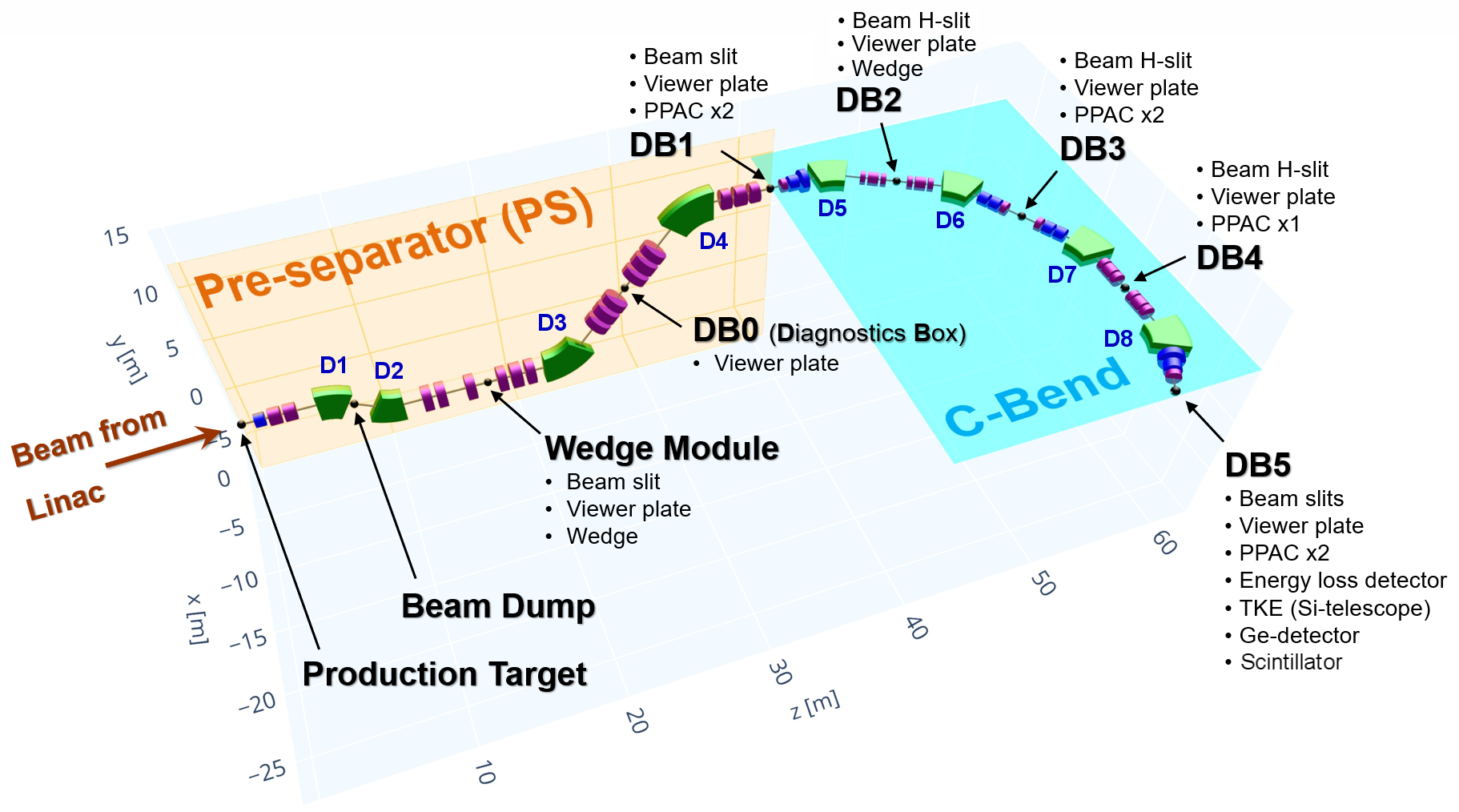}
    \end{minipage}
    \hfill
    \begin{minipage}{0.22\textwidth}
        \caption{Schematic view of the Advanced Rare Isotope Separator. Wedge materials were used in the pre-separator and diagnostic box two (DB2) of the C-Bend (see text). Position-sensitive PPACs were located at DB3  and DB4 (Part A) and DB3, DB4, and DB5 (Part B).}
        \label{fig:setup}
    \end{minipage}
\end{figure*}

The quest to stake out the limits of nuclear existence enjoys a broad, concerted experimental effort bolstered by theoretical models and predictions that serve as guideposts on the drive to driplines. Facilities like FRIB, FAIR, GANIL, and RIKEN will continue to push into the as-yet-unexplored neutron-rich region of the nuclear landscape, which promises to contain many exciting discoveries \cite{60Ca,Cortes2020,Santamaria2015,Li2023}. Ref. \cite{RIBWorld} provides a review of existing and future radioactive ion beam facilities and their capabilities.

Among the many experimental endeavors these facilities will undertake, the determination of the neutron dripline is one of the most fundamental. Mapping and modeling the neutron dripline is challenging due to the complex interplay of phenomena that determine whether a system is bound or not. The neutron dripline is currently known only to neon ($Z=10$) \cite{39Na_drip,39Na-22}. One aspect of FRIB's discovery potential is to establish the dripline for elements as high as molybdenum ($Z=42$) \cite{Neuf_drip}. The isotope discovery potential of FRIB has already been demonstrated in recent papers \cite{198Pt_FRIB,82Se_FRIB, FRIB10kW_U}.

After the experimental observation of a new isotope, one of the first basic questions that arises is that of the mass of the new species (as well as quantities derived therefrom). The determination of the mass informs us of important properties, such as deformation, shell evolution, and nucleon separation energies \cite{Mass_N_20-28,Mass_N_70,Mass_Yb}. Of particular interest and importance to neutron- (proton-)rich rare-isotope production near the driplines is the neutron (proton) separation energy. Production cross sections near the limits of existence are sensitive to this separation energy; weakly bound isotopes experience cross section suppression, compared with their more strongly bound neighbors.

Nuclear masses have implications far beyond rare-isotope production. The nuclear mass, and, in particular, the nuclear binding energy, are some of the most fundamental quantities of a nucleus and are vital components for nuclear decay and reaction rate calculations. The masses of many neutron-rich nuclei, especially near the neutron dripline, remain experimentally unknown. This necessitates the use of theoretical models and extrapolations to predict the masses of exotic nuclides. These mass models arise either from macroscopic observations about the nucleus (e.g., binding energy per nucleon, as in the classic liquid drop model \cite{LiquidDrop}), microscopic calculations of the nucleus derived from quantum mechanical principles \cite{Gogny_HFB, Neuf_drip, HFB22, HFB27, SVmin, NL3, UNEDF1}, a combination of both \cite{FRDM12, WS4RBF}, or an empirical parameterization of existing data \cite{KTUY-PTP05, DZ10}. The dozens of different mass models across all types are crafted to recreate the experimentally determined binding energies and their systematics (or, in the case of empirical parameterization, are directly derived therefrom). However, extrapolations of nuclear binding energies from these models yield disparate predictions. The poor agreement between these extrapolations has implications for both the derived quantities (such as the single-nucleon separation energy) and calculations that rely heavily upon them. In particular, astrophysical r-process abundance simulations depend greatly on theoretical extrapolations for both binding energies and reaction rates far from stability \cite{rProcess_reviewCowan}. Much attention has been given recently to the role of nuclear mass uncertainties in the r-process abundance calculations \cite{Mumpower_mass,Mumpower_properties,Mumpower_uncertainty}. The sensitivity of production cross sections of neutron-rich nuclei provides another test of nuclear mass predictions. Near the neutron dripline, small differences in binding energies, particularly neutron separation energies, can be amplified into large differences in predicted production cross sections. Thus, experimental cross sections offer an indirect probe of nuclear binding far from stability, although their interpretation also depends on the underlying reaction mechanisms.

In this work, we present newly measured production cross sections from the projectile fragmentation of an $^{82}$Se beam, as well as the first observation of the isotope $^{71}$Cr, which was separated with the Advanced Rare Isotope Separator (ARIS) \cite{ARIS2023} at the Facility for Rare Isotope Beams (FRIB) \cite{FRIB_commission}. These cross sections, in addition to those previously reported \cite{82Se_FRIB, MSU13}, are analyzed with the recently developed $\Delta BE$ systematics \cite{82Se_FRIB}. These systematics quantify the production cross sections in terms of the binding energy and, in particular, the one-neutron separation energy. Cross section predictions of unobserved nuclei are calculated from these systematics as well as from abrasion-ablation calculations. The uncertainty of these predictions is quantified by a Bayesian-inspired model averaging based on eleven mass models of various types.


\section{Experiment}

\begin{figure*} [htbp!]
    \centering
    \begin{minipage}{0.75\textwidth} 
        \includegraphics[width=\textwidth]{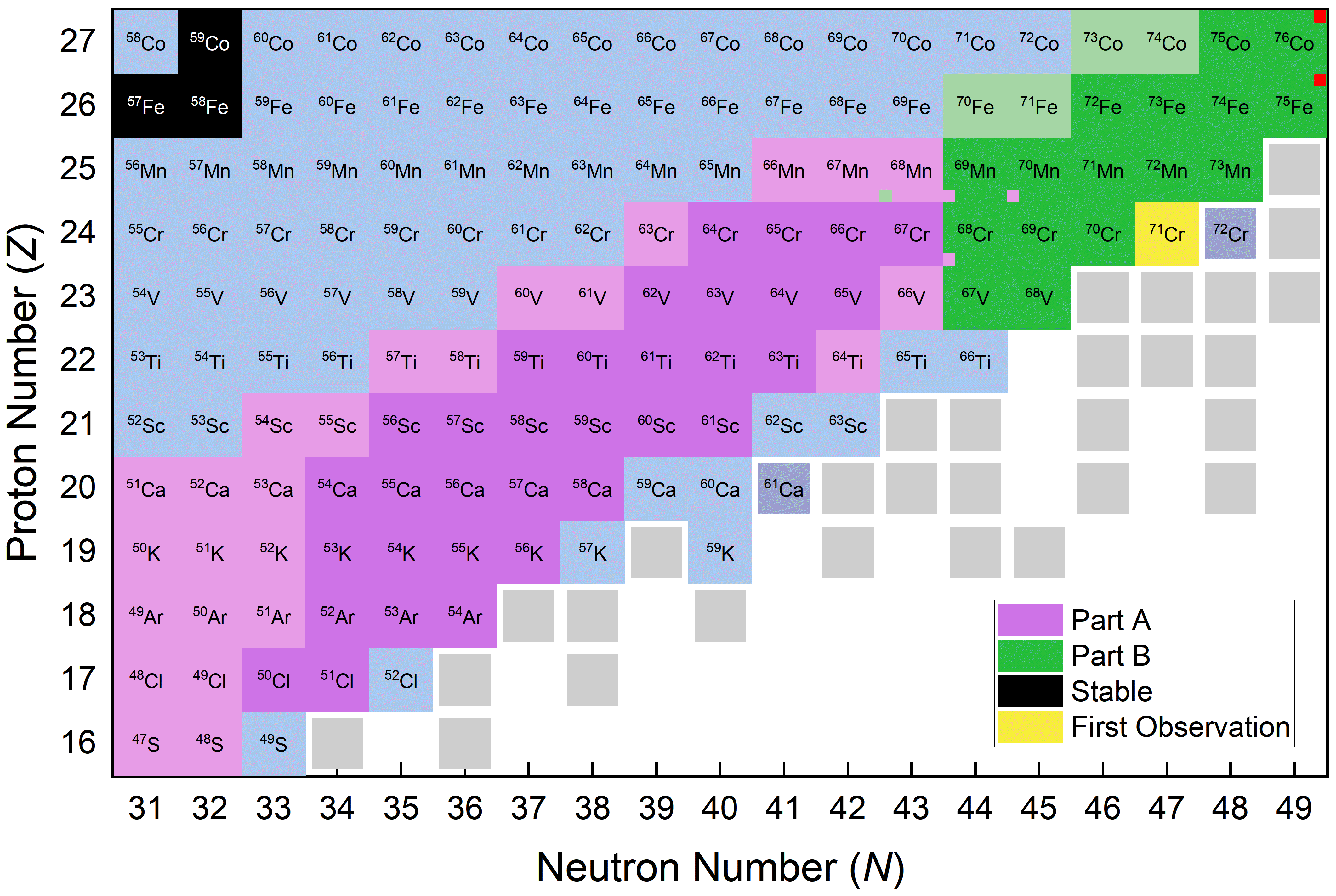}
    \end{minipage}
    \hfill
    \begin{minipage}{0.22\textwidth}
        \caption{Region of the chart of nuclides relevant to the experiments conducted. The primary fragments produced and studied in Part A are colored purple, while the primary fragments produced in Part B are colored green. The darker shading of each color indicates which isotopes had production cross sections measured. The black squares denote stable isotopes. The newly observed isotope $^{71}$Cr is colored yellow. $^{61}$Ca is marked because of the potential single count observed in Ref.~\cite{82Se_FRIB}. $^{72}$Cr was recently observed from the fission of uranium~\cite{72Cr}. The smaller gray squares represent isotopes that are predicted to exist by the HFB-22 mass model, but have yet to be observed. $^{75}$Fe and $^{76}$Co (marked by the red square in the corner) are from net neutron pick-up reactions.}
        \label{fig:NuclideChart}
    \end{minipage}
\end{figure*}

The production cross sections were measured during two separate but complementary experiments at FRIB. These data sets are denoted as \textit{Parts A} and \textit{B}. In both parts, an \nuc{82}{Se} beam was accelerated by the FRIB linear accelerator and fragmented on a carbon target at the entrance of ARIS. The resulting rare isotopes were transported to either the FRIB Decay Station initiator (FDSi, Part A) or the fifth diagnostic box (Part B). Figure \ref{fig:setup} illustrates the ARIS system and denotes the location of detectors and materials along the beamline. The region of interest in these complementary experiments was the neutron-rich chlorine to cobalt region, $17\leq Z \leq 27,\,\, 33\leq N \leq 49$. Figure \ref{fig:NuclideChart} depicts the regions covered in each experiment. The purple and green boxes denote primary fragments observed in Parts A and B, respectively. The darker shading of each color denotes isotopes for which production cross sections were measured. The small purple and green squares in the lower left corners of the chromium and manganese isotopes indicate that those isotopes were observed in both experiments. The small red squares in the upper right corners of the $N=49$ iron and cobalt isotopes indicate net neutron pick-up reactions.

The similarities and differences in the experimental setups are discussed in the following subsections.

\subsection{Part A}

The first data set was collected during an experiment involving the FDSi \cite{FDSiProp, FDSi_NPN}. The experimental setup of the decay station is detailed in recently published work \cite{Crawford2022,Gray2023FDSi,Peltier2025FDSi,Cox2024FDSi}, including other analyses of data from this experiment \cite{e21069B_Ong, Xu_54K}. 

The FDSi experiment utilized a primary beam of 215 MeV/$u$ $^{82}$Se, which was impinged onto a 5 mm-thick rotating $^{12}$C target ($\rho=1.89$ g/cm$^3$). The primary beam intensity had an average value of 0.543 particle $\mu A$. There were two spectrometer settings corresponding to the two fragments of interest: $^{55}$Ca (with magnetic rigidity B$\rho$ = 5.2642 Tm) and $^{54}$K (with magnetic rigidity $B\rho$ = 5.3613 Tm). The production cross sections were measured during the $^{54}$K runs.

The $^{54}$K production setting used a 1.5 mm-thick momentum-compressing \cite{BANDURA2011} aluminum-7075 alloy wedge ($\rho = 2.81$~g/cm$^3$) at the focal plane of the vertical pre-separator \cite{kahlWedge}. A 1 mm-thick aluminum wedge of the same alloy was located at the second diagnostic box (DB2) of the horizontal C-Bend. Particle identification (PID) was performed event-by-event with the $B\rho$-$\Delta E$-ToF method. This method is outlined in the appendix of Ref.~\cite{76Ge_PRC}.

\begin{figure}
    \centering
    \includegraphics[width=1.0\linewidth]{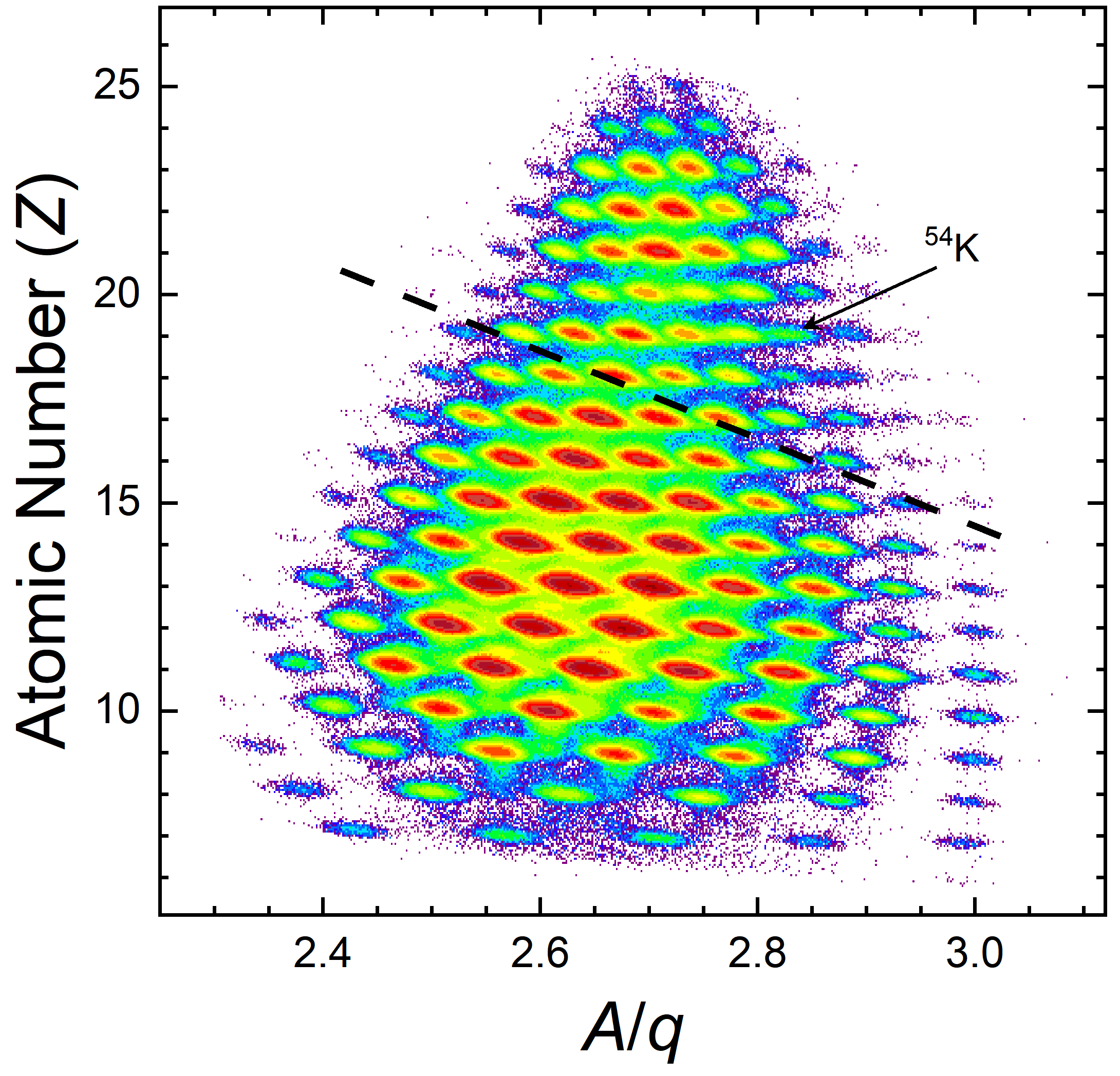}
    \caption{Particle identification for Part A, shown as a function of mass-to-charge ratio $A/q$ and atomic number $Z$. \nuc{54}{K} is shown for reference. Fragments below the black dashed line were dominated by interactions in the wedge or transmitted through high-order aberrations in the system. The location of the line was determined by a \lisepp simulation.}
    \label{fig:e21069B_PID}
\end{figure}

The magnetic rigidity ($B\rho$) was measured with an NMR probe in the last dipole of ARIS. Corrections to $B\rho$ were made event-by-event using measured position data from delay-line position-sensitive parallel-plate avalanche counter (PPAC) \cite{PPAC} located at DB4 (dispersive focus). This correction to $B\rho$ produces an enhanced particle identification resolution through the use of trajectory reconstruction techniques \cite{HIAT25, BerzReconstruction}, which provide a more accurate determination of momentum deviation ($\delta p/p$). The time-of-flight (ToF) was measured between the second PPAC at DB3 and the fast plastic scintillator at the FDSi first focal plane (FP1). Two independent ToF measurements, obtained using the PPAC up and down channels, were averaged to improve the timing resolution. Energy loss ($\Delta E)$ was measured at FP1 with the MSX100 silicon PIN diode detector.


The measurements of magnetic rigidity, time-of-flight, and energy loss are used to calculate the atomic number ($Z$) and the mass-to-charge ratio ($A/q$) of the observed fragments. All of the observed fragments were fully stripped and no charge-state contamination was seen. This leads naturally to a particle identification plot, as shown in Figure~\ref{fig:e21069B_PID}. The black dashed line separates primary fragments, which are produced mainly in the production target, from secondary fragments, which are dominated by reactions in the wedge materials (effectively behaving as a secondary production target) or are transmitted via high-order aberrations in the optical system, regardless of origin. The primary fragments, being above the line, are characterized by high transmission efficiency in first-order optics calculations.

The location of this boundary was determined by \liseppcute \cite{LISE23} (v.18.2) simulations. In particular, the high-order aberration contributions were studied with Monte Carlo simulations of ARIS using fifth-order COSY INFINITY \cite{COSY} maps in \liseppsh. Additional calculations were also performed in which the pre-separator wedge behaved as a secondary production target. Isotopes whose simulated production was significantly enhanced by the wedge material were placed below the line.

Isotopes above the dividing line are predominantly produced by fragmentation in the primary target, although contributions from reactions in the wedge materials and transmission through high-order optical aberrations may still be present. Below the line, these additional contributions become more significant and may dominate the observed yields.

An additional criterion for cross section measurement is the overall transmission of an individual isotope. Nuclides with a small transmission efficiency in first-order optics subsequently experience large transmission changes for small deviations in optical parameters. Transmission parameters and uncertainty quantification are discussed further in Section~\ref{sec:Results}. For the analysis conducted here, a transmission greater than 0.1\% was required in the baseline optics setting.

\begin{figure*} [htbp!]
    \centering
    \begin{minipage}{0.75\textwidth} 
        \includegraphics[width=\textwidth]{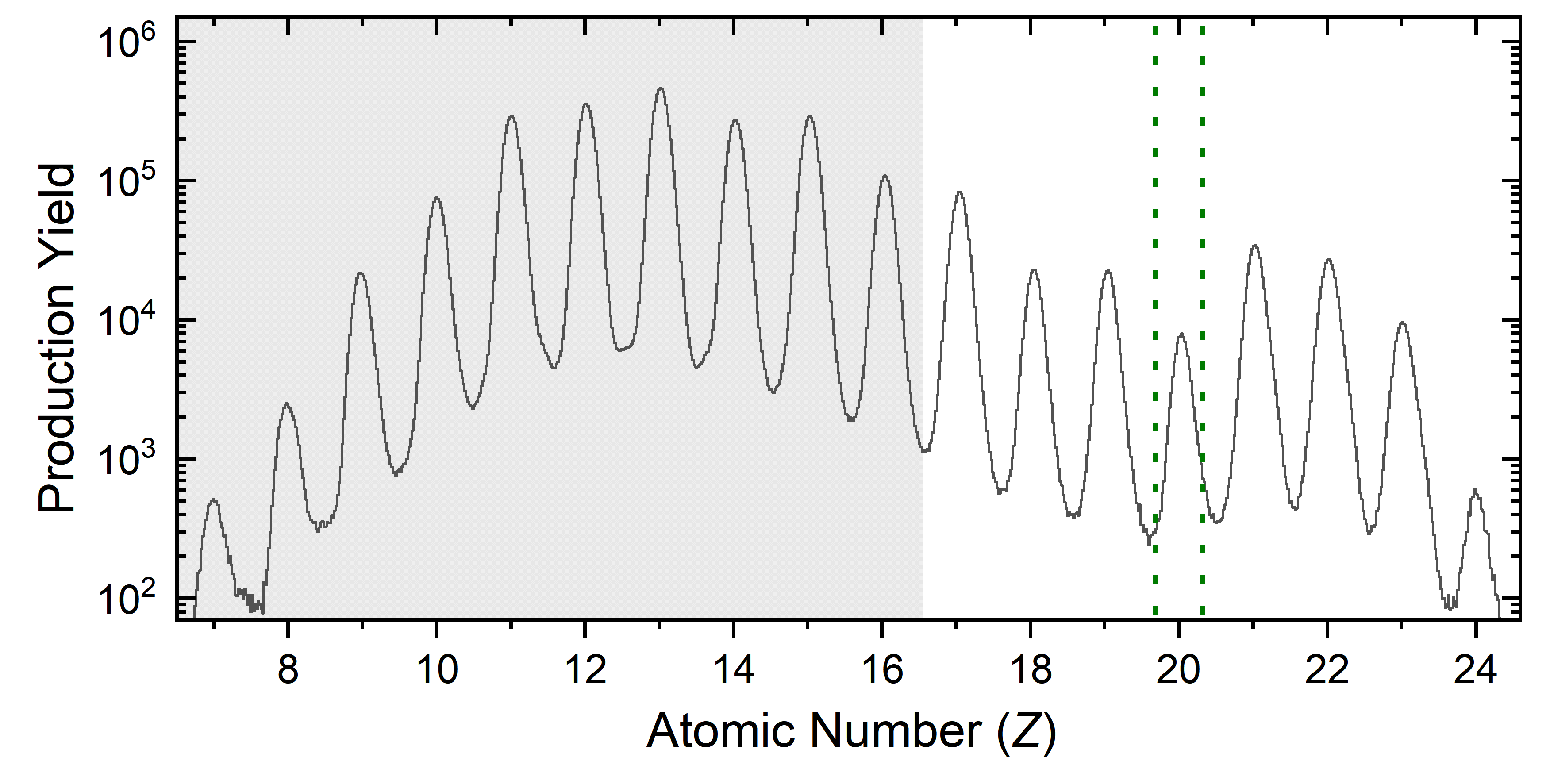}
    \end{minipage}
    \hfill
    \begin{minipage}{0.22\textwidth}
        \caption{Elemental distribution from Part A.  The shaded region denotes elements that were not analyzed for production cross sections (dominated by wedge products and high-order aberrations). Gaussian functions of a common width were fit to the distribution to determine the standard deviation of the elemental peaks. The dotted green line represents the gate range corresponding to $\pm3 \sigma$ used to distinguish calcium. The projection of the calcium isotopes is shown in Fig.~\ref{fig:Masses}(a). The $Z$ resolution was determined to be $\sigma(Z)=0.11$.}
        \label{fig:Z21069B}
    \end{minipage}
\end{figure*}

The elemental spectrum from Part A is shown in Figure~\ref{fig:Z21069B}. The peaks of the spectrum were fitted with Gaussian functions constrained to a common sigma. The dashed green line shows the gate range that was used to determine the extent of the calcium isotopes. This range was chosen to correspond to three standard deviations, $\pm\,3\sigma$, of the fitted distribution. The $Z$ resolution was determined to be $\sigma(Z)=0.11$. The elements within the shaded gray region were dominated by reactions in wedge materials and their isotopes did not have cross sections measured. Across the entire range shown ($Z=7-24$), there is consistent elemental identification. The energy loss detectors were well calibrated and well suited for such a broad range of fragments.

\begin{figure*} [htbp!]
    \centering
    \begin{minipage}{0.75\textwidth} 
        \includegraphics[width=\textwidth]{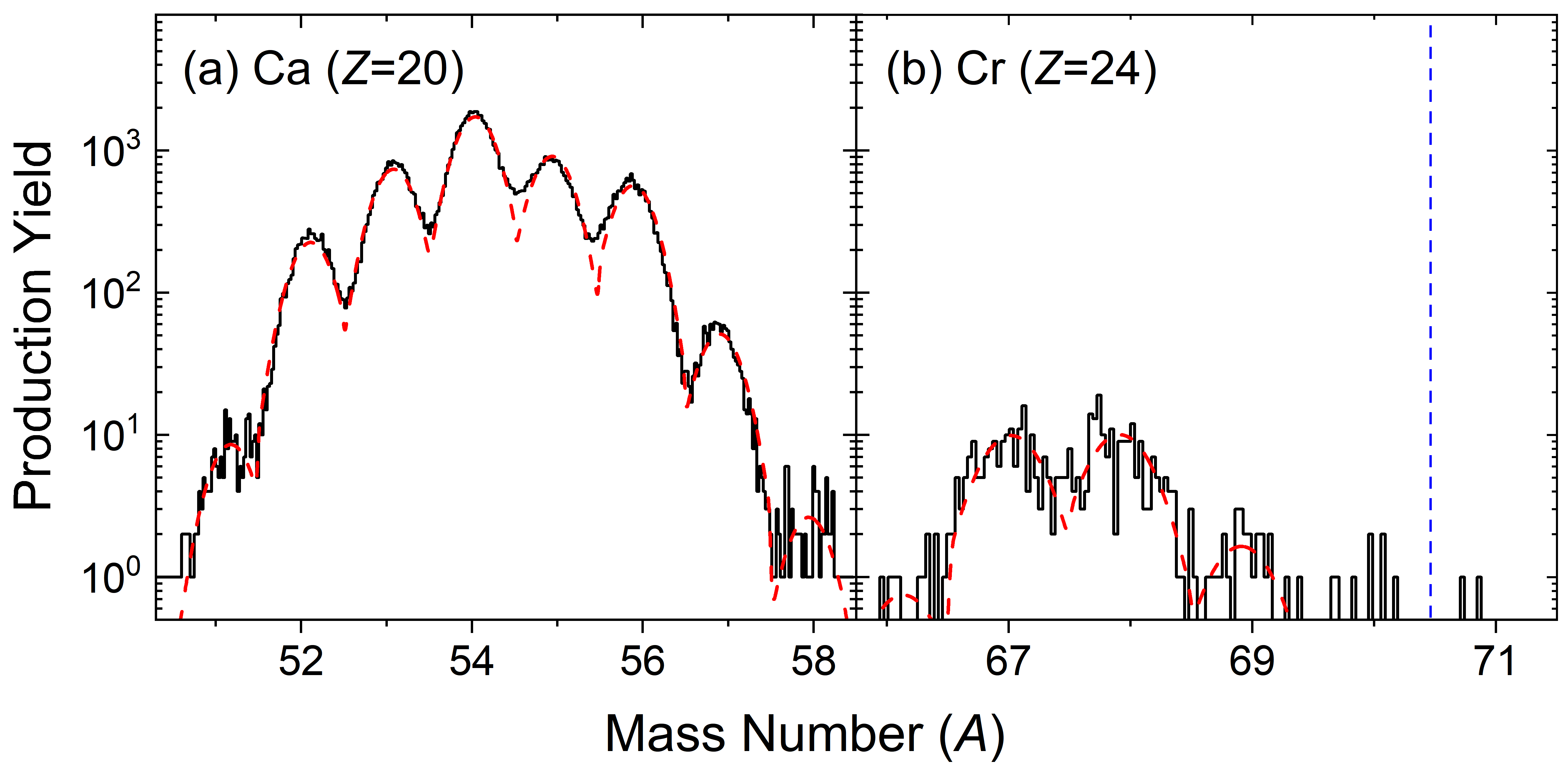}
    \end{minipage}
    \hfill
    \begin{minipage}{0.22\textwidth}
        \caption{Isotopic distributions for calcium (Part A) and chromium (Part B). Gaussian functions of a common width were fitted to each isotopic spectrum. The isotopic yield was determined by integrating the fitted Gaussian peaks. The mass resolution for Part A was determined to be $\sigma(A)=0.24$. The mass resolution for Part B was determined to be $\sigma(A)=0.24$. Counts to the right of the dashed blue line in panel (b) are the newly observed $^{71}$Cr.}
        \label{fig:Masses}
    \end{minipage}
\end{figure*}

The mass resolution, determined to be $\sigma(A)=0.24$, is displayed for calcium in panel (a) of Figure~\ref{fig:Masses}. Note that the dotted red lines represent the fitted Gaussian distribution of constant width from which the isotopic yield statistics were derived.

\subsection{Part B}

The second data set was collected during beamline commissioning to the S1 vault, the new home of the FDSi~\cite{FDSiMove}. Here again, a primary beam of $^{82}$Se, this time at 225 MeV/$u$ with an average intensity of 1.0 particle $\mu A$, was impinged onto an 8 mm-thick rotating $^{12}$C target ($\rho = 1.89$~g/cm$^3$). The experiment had two goals: commission the beamline to the S1 vault and perform a new isotope search for $^{71}$Cr.

\begin{figure}
    \centering
    \includegraphics[width=1.0\linewidth]{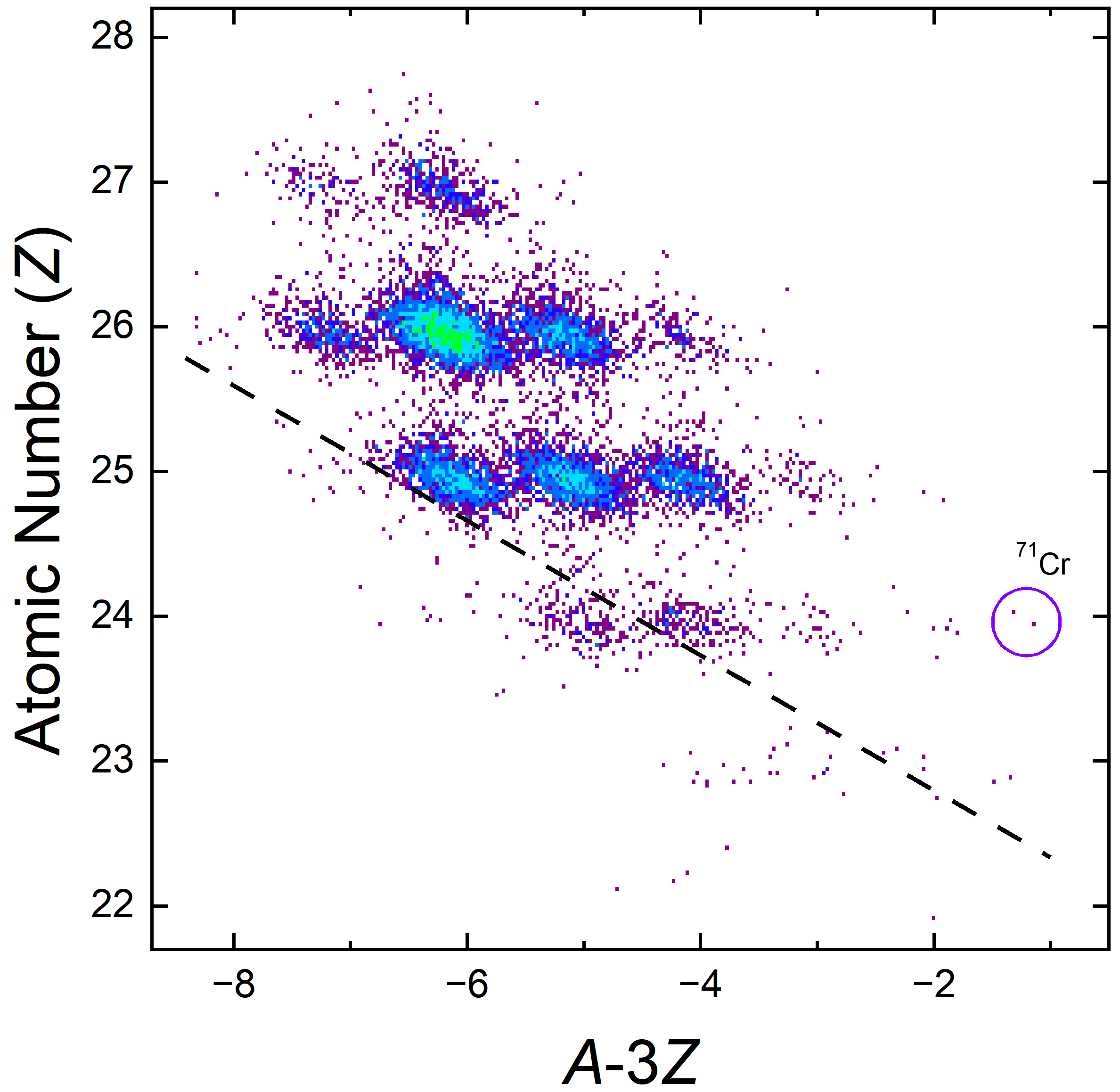}
    \caption{Particle identification for Part B, shown as $Z$ vs $A-3Z$. Here, $A-3Z$ is calculated as $(A/q-3)\times Z$, where only the fully stripped charge state was selected. Fragments below the black dashed line were created by interactions in the wedge or transmitted by high-order aberrations in the system. The two counts associated with the newly identified isotope, $\nuc{71}{Cr}$, are highlighted by the purple circle.}
    \label{fig:71Cr_PID}
\end{figure}

The new isotope production runs used a 1.2 mm-thick pre-separator momentum-compressing aluminum-6061 alloy wedge ($\rho = 2.7$~g/cm$^3$) and a 2 mm-thick DB2 wedge of the same alloy. Particle identification was again performed event-by-event with the $B\rho$-$\Delta E$-ToF method. The time-of-flight was measured between PPACs at DB3 and DB5. Two timing measurements were averaged to obtain improved timing resolution. Energy loss and total kinetic energy (TKE) were measured by a PIN diode detector (500 $\mu$m thick) and a silicon telescope with six layers of varying thickness at DB5 (1000, 1000, 1000, 1500, 1500, and 1500 $\mu$m, respectively). 
The additional measurement of TKE allows for the unambiguous identification of charge states (for which there was some contamination from hydrogen-like nuclei) and permits the construction of a particle identification plot with $A-3Z$, as seen in Figure~\ref{fig:71Cr_PID}. The calibration of the PID in Part B was confirmed by the strong 908 keV isomeric gamma line in \nuc{78}{Zn} during a different beam setting. The energy-loss, TKE, and time-of-flight calibrations were consistent between runs, enabling unambiguous identification of the neutron-rich isotopes produced in the \nuc{71}{Cr} setting.

\begin{figure}
    \centering
    \includegraphics[width=1.0\linewidth]{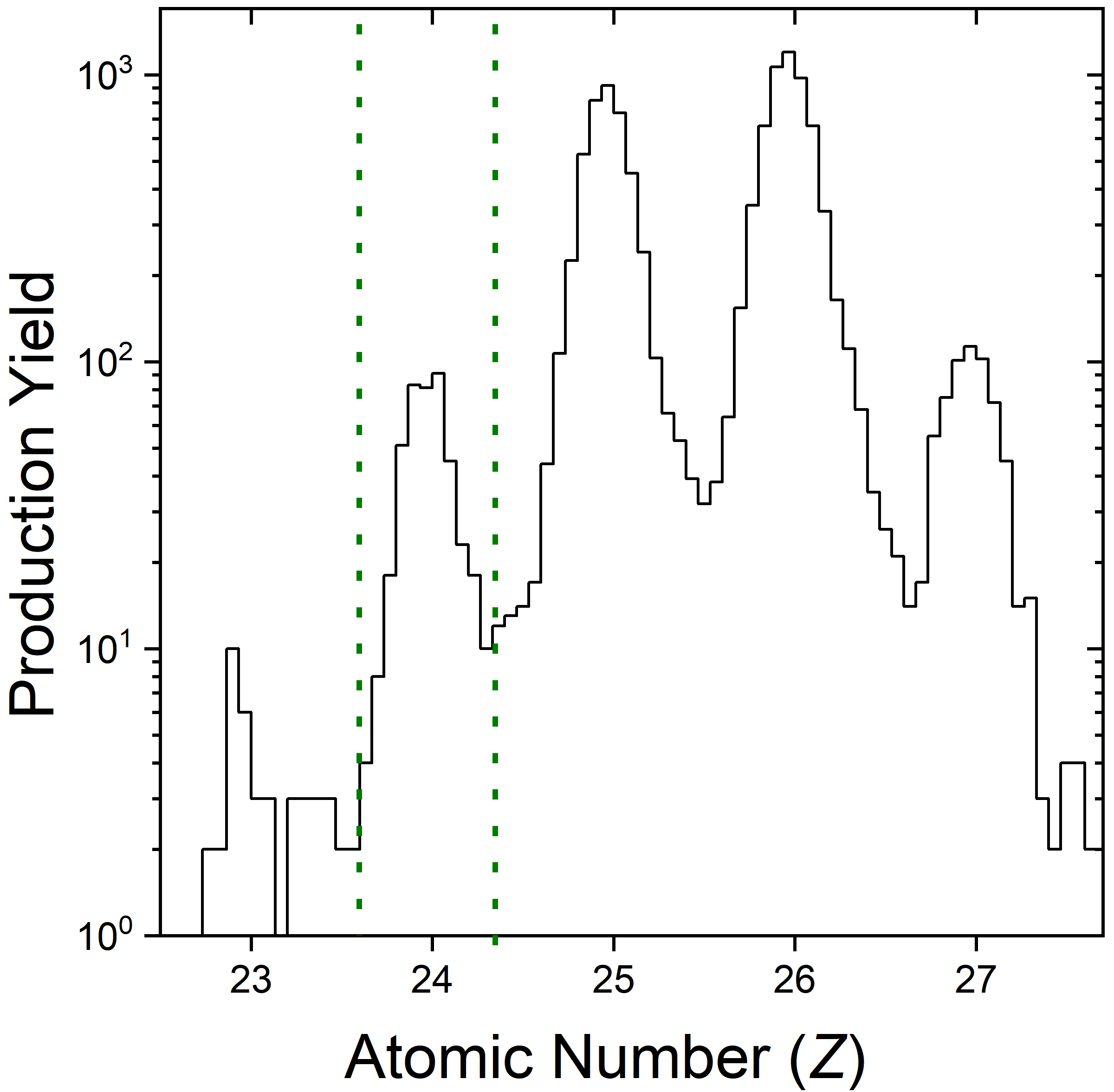}
    \caption{Elemental distribution from Part B. The dotted green line represents the gate range corresponding to $\pm3 \sigma$ used to distinguish chromium. The projection of chromium isotopes is shown in Fig.~\ref{fig:Masses}. The $Z$ resolution was determined to be $\sigma(Z)=0.13$.}
    \label{fig:71Cr_Z}
\end{figure}

As in Part A, events below the black dashed line are primarily attributed to reactions in the wedge materials or to transmission through high-order ion-optical aberrations. The location of this boundary was determined through \lisepp transmission and wedge reaction simulations. 

The elemental spectrum from Part B is shown in Figure~\ref{fig:71Cr_Z}. As in Part A, the peaks of the distribution were fit with Gaussian functions constrained to a common width. The dashed green line shows the gate range corresponding to $\pm3\sigma$, which was used to determine the extent of chromium. The $Z$ resolution was determined to be $\sigma(Z)=0.13$. 

The mass resolution of Part B is shown in panel (b) of Figure~\ref{fig:Masses}, which was determined to be $\sigma(A)=0.24$. The red dashed line represents the fitted Gaussian distribution from which the isotopic yield statistics were derived. The two counts to the right of the dashed blue line are associated with the newly observed $^{71}$Cr.

The resolution of the particle identification quantities ($\sigma(A), \sigma(Z)$), along with suppression of the background, the method for which is outlined in Appendix A of Ref.~\cite{82Se_FRIB}, allows for a confident and unambiguous assignment of the new isotope counts. The probability that a single event arises from the tail of a more intense neighboring isotope peak is approximately $1\times 10^{-7}$. 

\section{Results}
\label{sec:Results}

The production cross sections were deduced from the experimental data through the relationship
\begin{equation}
    \sigma=\frac{Y}{\xi\, B_{\mathrm{int}}\, t},
\end{equation}
where $Y$ is the number of detected events for a given nuclide (fragment yield), $\xi$ is the transmission efficiency calculated using \lisepp simulations, $B_{\mathrm{int}}$ is the integrated number of beam particles incident on the target, and $t$ is the target areal density in atoms/cm$^2$, calculated from the target density and thickness using Avogadro's constant, $N_A$.

Fragment yields were determined by fitting the mass distributions with Gaussian functions of a common width and integrating the fitted peaks, as illustrated in Fig.~\ref{fig:Masses}(a) and (b). 
Statistical uncertainties were evaluated using Poisson counting
statistics, with the low-count treatment following
Ref.~\cite{countingStats}.

The transmission efficiency $\xi$ was determined using \lisepp simulations.
The simulated fragment distributions were matched to the experimental data
by varying the target and wedge thicknesses, their thickness nonuniformities
(referred to as ``thickness defects'' in \liseppsh), slit positions, and the
vertical offset of the pre-separator wedge.
The resulting parameter set was defined as the ``nominal setting''. The estimated transmission uncertainty $\Delta \xi$ was determined by varying the parameters slightly above and below the nominal setting values, following the procedure outlined in Ref.~\cite{76Ge_NIMA}. The final transmission was taken to be the median of all resultant transmission points, with the uncertainty bounds set according to the minimum and maximum transmission deviation. The non-linear transmission response of individual isotope species to the varied parameters produces asymmetric transmission errors. For isotopes with small simulated transmission values, this is the largest source of uncertainty in the cross-section measurements.

The integrated number of beam particles was determined from the power measurements of a non-intercepting probe located in front of the target. The error in the measurements was estimated to be 10\% in both experiments. The beam power, when converted to the number of particles per second (pps) based on the primary beam energy, was, on average, $3.39\times 10^{12}$ pps (Part A) and $6.43\times 10^{12}$ pps (Part B).

The target thickness uncertainty in both experiments was determined by the difference between the nominal thickness and the measured thickness. This corresponded to 3.75\% and 2.40\% for Parts A and B, respectively. This value is reflected in the cross section uncertainty. The target defect was taken to be 2\% and 1\% in Parts A and B, respectively, and is included in the transmission calculations in \liseppsh. The target areal densities were $4.74 \times 10^{22}$ atoms/cm$^2$ (Part A) and $7.58\times 10^{22}$ atoms/cm$^2$ (Part B).

The reported cross section uncertainties are the result of adding the contributions described above in quadrature. The upper and lower uncertainties are calculated separately to preserve their asymmetric nature.

The measured cross sections, shown as the dark purple and green squares in Figure~\ref{fig:NuclideChart}, were combined with previously measured data from Refs.~\cite{MSU13,82Se_FRIB}. For isotopes with multiple measurements, the cross sections were combined using a logarithmic averaging procedure that preserves the asymmetry of the uncertainties. The previously reported production cross sections are treated as independent measurements of the true production cross section. Although the production mechanism is the same (\nuc{82}{Se} fragmentation), there are slight variations in the target thicknesses, primary beam energies, and optics across the different measurements. In averaging the values, the widths of the error bars were reduced when the central values agreed.

\begin{figure*}
    \centering
    \begin{minipage}{1.0\textwidth}
    \includegraphics[width=0.8\linewidth]{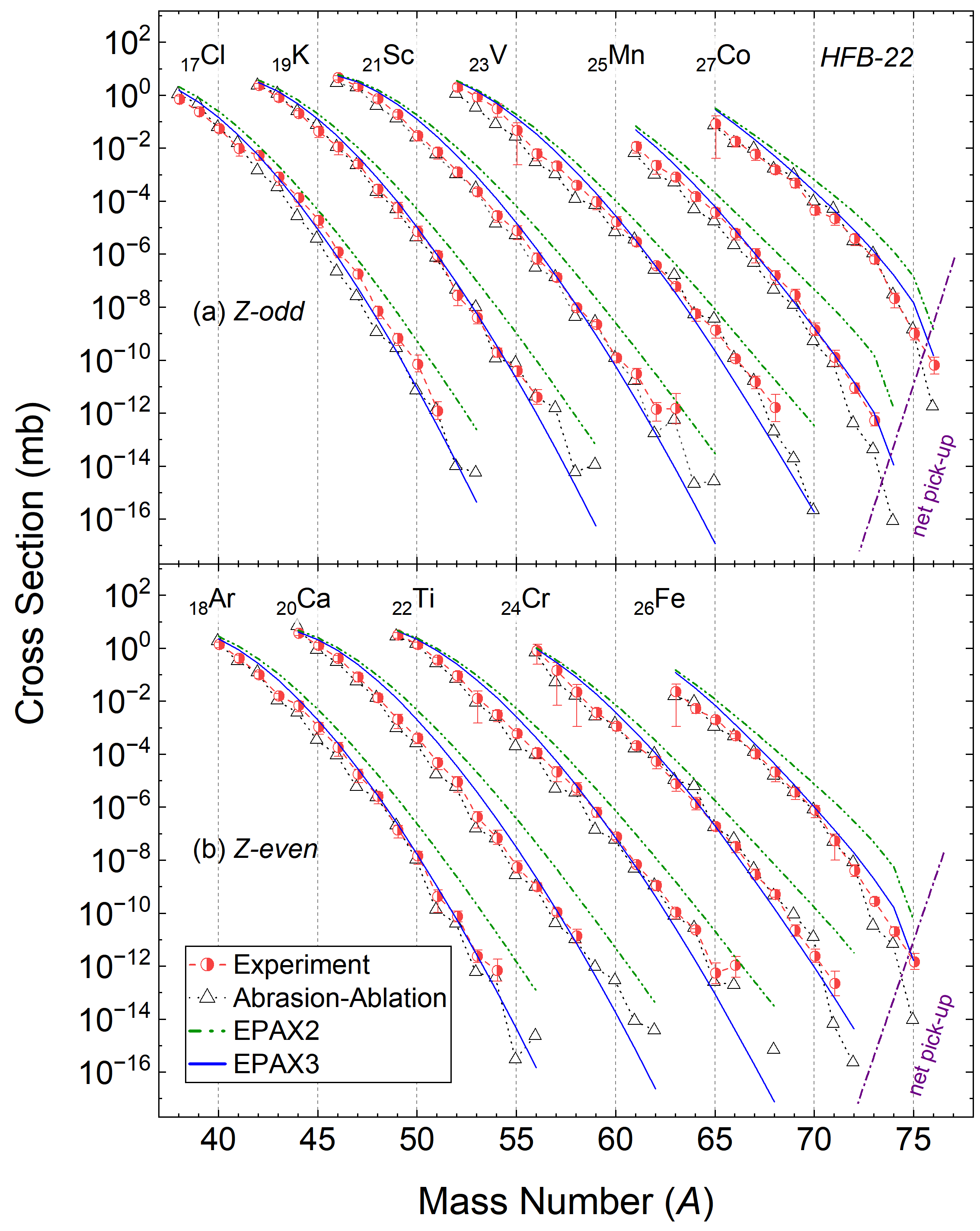}
    \caption{Comparison of experimentally measured cross sections from this work and from Refs.~\cite{MSU13,82Se_FRIB} (red half circles), abrasion-ablation calculations (open black triangles) based on the HFB-22 mass table, and EPAX2~\cite{EPAX2} and EPAX3~\cite{EPAX3} predictions (green and blue lines, respectively). Isotopes resulting from net neutron pick-up are denoted by the dark purple dot-dash line.}
    \label{fig:CS_AA_EPAX3}
    \end{minipage}
\end{figure*}

The production cross sections are plotted in Figure~\ref{fig:CS_AA_EPAX3}, together with the EPAX2 and EPAX3 parameterizations \cite{EPAX2, EPAX3}, and results from the abrasion-ablation calculations based on the HFB-22 mass table, which are discussed in the next section.

\subsection{Abrasion-Ablation}

The abrasion-ablation (AA) calculations were performed with \lisepp \cite{LISE_AA}. Because of the breadth of the cross section data, the data set was broken up into three regions: low-$Z$ ($Z=17-19$), medium-$Z$ ($Z=20-24$), and high-$Z$ ($Z=25-27$). The regions were chosen to provide an accurate description of the experimental cross sections and are  roughly connected to the shell gaps at $Z=20, 28$. 

The abrasion-ablation model takes into account the excitation energy generated during the abrasion stage and the statistical particle evaporation during the ablation stage. The excitation energy model used in these fits takes the form of an exponential distribution \cite{meanT_dist}, whose scale is set by the effective excitation energy scale parameter $T_{\mathrm{eff}}$ of the nucleus following abrasion:

\begin{equation}
    T_{\mathrm{eff}}=c_1+c_2\Delta A_{\mathrm{abr}}+c_{NZ}\left(\Delta N_{\mathrm{abr}}-\Delta Z_{\mathrm{abr}}\right),
    \label{eq:eE_calc}
\end{equation}
where $\Delta A_{\mathrm{abr}}$, $\Delta N_{\mathrm{abr}}$, and $\Delta Z_{\mathrm{abr}}$ denote the numbers of nucleons, neutrons, and protons abraded from the projectile nucleus, respectively. The coefficients $c_1$, $c_2$, and $c_{NZ}$ were fitted parameters in the AA minimization. Within this parameterization, $T_{\mathrm{eff}}$ controls the excitation-energy distribution and therefore the subsequent particle evaporation. A higher initial excitation energy generally leads to greater particle evaporation and, where neutron evaporation dominates, shifts the predicted yield toward lower mass numbers along a given isotopic chain.

Additional parameters fitted in the AA calculations include the effective Coulomb barrier, which modifies the probability of charged-particle evaporation; the odd-even staggering, which is a phenomenological parameter that is used to reproduce the pairing correlation effect observed near the driplines; and break-up temperatures, which set the scale of excitation energy where the nucleus undergoes complete disassembly into its constituent nucleons and light clusters. An overall cross-section scaling factor was also determined. The fitted parameters determined in this work are compiled in Table~\ref{tab:AAcoef} in Appendix A. 

The deduced parameter from the AA minimization $\mathcal{J}_{\mathrm{AA}}$ for eleven mass models is tabulated in Table~\ref{tab:comparison_AA}.  $\mathcal{J}_{\mathrm{AA}}$ is the objective function and is the sum of two contributions: a quadratic component and a logarithmic component. The former emphasizes agreement in the absolute value of the cross sections, while the latter increases the sensitivity of the minimization to the small cross sections of the most exotic isotopes. The relative weights assigned to these components were chosen such that they contributed approximately equally to the overall deduced parameter. The functional forms of these components are given in Ref.~\cite{82Se_FRIB}. The deduced parameter also incorporates a penalty function when a mass model predicts that an observed isotope is unbound. The penalty treats the unbound nucleus as under-predicting the measurement by a factor of ten. The penalty is incorporated into both the quadratic and logarithmic components.

The abrasion-ablation cross sections plotted in Figure~\ref{fig:CS_AA_EPAX3} are based on the HFB-22 mass table. Among the considered mass models listed in Table~\ref{tab:comparison_AA}, HFB-22 provides a broadly good description across the three $Z$ regions and is used for the representative AA calculations shown. Overall, there is good agreement between the calculations and the experimental values. There is an under-prediction of the chlorine isotopes. The kink created at \nuc{55,56}{Ar} arises from the fact that HFB-22 predicts \nuc{55}{Ar} to be ``just bound'' with a neutron separation energy of 0.05 MeV. Other mass models predict it to be unbound.


The results of the AA calculations for $^{63}$Sc$_{42}$ and $^{66}$Ti$_{44}$ are particularly noteworthy. In these isotopic chains, the most neutron-rich nuclide experiences significant enhancement compared to what is expected based on the exponential cross-section trend. The calculated cross sections are able to qualitatively recreate these kinks. However, the magnitude of enhancement for \nuc{66}{Ti} is lower by an order of magnitude. The inability of the AA calculations to quantitatively reproduce these kinks suggests that some processes not fully included in the present model, such as multistep reactions in the primary target, may contribute to the unusually large measured cross sections. This possibility will be discussed further in the context of the $\Delta BE$ systematics.


In the heaviest elements studied (chromium-cobalt), there is a significant under-prediction of the most neutron-rich isotopes. This is potentially a result of contributions from charge-exchange reactions during fragmentation. These processes have been studied both experimentally \cite{expCE} and theoretically \cite{theoryCE}.

The \lisepp package includes a phenomenological treatment of charge exchange in abrasion-ablation calculations. This extension was motivated by the observation of fragments containing more neutrons than the incident $^{82}$Se projectile, both in the present experiment and in previous measurements (see, e.g., Ref.~\cite{MSU13,60Ca}). The charge-exchange parameterization was guided by results from BeAGLE simulations~\cite{Ddamulira_2026}. BeAGLE is a multistage Monte Carlo event generator~\cite{BeAGLE} that incorporates intranuclear-cascade (INC) dynamics and subsequent statistical de-excitation.

In the current beta implementation of \liseppsh, charge exchange is introduced following the abrasion stage, allowing the conversion of one nucleon species into the other ($n \rightarrow p$ or $p \rightarrow n$). Pure nucleon pick-up is not included. After determining the numbers of abraded protons and neutrons, the charge-exchange cross section is evaluated using a phenomenological parameterization scaled to the geometrical cross section of the projectile-target system. The resulting pre-fragment composition is passed to the excitation-energy calculation, followed by statistical de-excitation.

Including charge exchange significantly enhances the calculated production cross sections of neutron-rich fragments near the pure-proton-removal limit, particularly for nuclei with $N=48$. It also enables the calculation of production cross sections for the experimentally observed $N=49$ fragments, which cannot be produced through neutron removal alone. However, the present implementation is limited to single-nucleon charge exchange. Contributions from multiple charge-exchange processes and direct neutron pick-up may also be important
and require further investigation.

\begin{table}[ht]
\setlength{\abovecaptionskip}{20pt} 
\centering
\setlength{\extrarowheight}{2pt}
\caption{Deduced parameter $\mathcal{J}_\mathrm{AA}$ from the minimization of abrasion-ablation calculations for \nuc{82}{Se} projectile-fragmentation products in the mass ranges $17 \leq Z \leq 19$ (Low), $20 \leq Z \leq 24$ (Medium), and $25 \leq Z \leq 27$ (High) and the deduced parameter $R_\mathrm{\Delta BE}$ for eleven mass models. Masses beyond the known range of AME2020~\cite{AME2020} were extrapolated using the LDM3 model implemented in \liseppsh.}

\begin{ruledtabular}
\begin{tabular}{l@{\hspace{0.10cm}}
                >{\raggedleft\arraybackslash}p{0.65cm}|
                >{\centering\arraybackslash}p{0.68cm}
                >{\centering\arraybackslash}p{1.08cm}
                >{\centering\arraybackslash}p{0.68cm}
                |l}
\multirow{2}{*}{\raisebox{-12pt} {\parbox[c]{1.2cm}{\centering \hspace*{25pt}Mass\\\hspace*{25pt}Model}}} & 

&

\multicolumn{3}{c|}{ $\mathcal{J}_{\mathrm{AA}}$} & \multirow{2}{*}{\raisebox{-12pt} {\parbox[c]{1.0cm}{\centering \hspace*{-0.50cm} $R_{\mathrm{\Delta BE}}$}}} \\[-3pt]
 & & \multicolumn{3}{c|} {\hspace*{-15pt}\rule{0.43\linewidth}{0.4pt}} \\
 & & Low & Medium & High & \\[3pt]
\hline
AME2020     & \cite{AME2020}   & 2.29  & 1.61 & 1.65   &  2.54 \\ 
FRDM2012       &  \cite{FRDM12}      & 1.69   & 1.71 & 1.23  & 1.92  \\ 
HFB-22     & \cite{HFB22}  & 0.96   & 1.00 & 1.33  &  0.93 \\ 
HFB-27     & \cite{HFB27}      & 1.52   & 1.53 & 1.78  &  1.05 \\ 
KTUY    & \cite{KTUY-PTP05}    & 1.58   & 1.59 & 1.87  &  1.31  \\ 
LISE-LDM3  & \cite{LISE08}     & 17.82   &  2.51 & 1.51  &  3.09 \\ 
SV-MIN     & \cite{SVmin}   & 1.04   & 1.50 & 1.67  & 2.14 \\ 
TUYY & \cite{TUYY}    & 2.23   & 2.18 & 1.33  &  3.24 \\ 
UNEDF0       & \cite{UNEDF0}   & 0.80   &  1.18 & 2.52  & 1.65 \\ 
UNEDF2       & \cite{UNEDF2}   & 1.45   &  2.82 & 1.59  & 6.10 \\ 
WS4$_{RBF}$     & \cite{WS4RBF}     & 4.15   & 2.67 & 1.55  & 3.27 \\ 
\end{tabular}
\end{ruledtabular}
\label{tab:comparison_AA}
\end{table}


The charge exchange option was used for all regions. Fragments that experienced larger net neutron removal were less influenced by the charge-exchange contribution. The dramatic under-prediction of \nuc{71}{Cr} and \nuc{68}{V} potentially points to more significant charge-exchange contributions for few net-neutron removal nuclei than what was considered here. This provides another source of potential enhancement for the kinked nuclei. However, the magnitude of enhancement is not likely to be fully explained by this contribution.



Despite the additional structure of the measured and AA-predicted production cross sections, the EPAX3 parameterization remains suitable as a rough approximation near the neutron dripline in the calcium region. We emphasize here, however, that the EPAX3 parameterization is blind to nuclear structure effects and nucleon separation energy. Therefore, for isotopes that are weakly bound, the actual cross section may differ by orders of magnitude in increasingly neutron-rich regions.


\subsection{\texorpdfstring{$\Delta BE$}{dBE} Systematics}
\label{dBEsys}

The $\Delta BE$ systematics \cite{82Se_FRIB} are defined by

\begin{equation}
    \Delta BE_{\mathrm{sys}} = k_1 \ln(\Delta BE_{\mathrm{max}}+1) + k_2 \ln(S_n) +b
    \label{eq:dBE}
\end{equation}
where $\Delta BE_{\mathrm{max}}$ is the difference in binding energy between the highest binding energy in the isotopic chain ($Z$, $N_{\mathrm{BEmax}}$) and the isotope under consideration ($Z$, $N$). $S_n$ is the one-neutron separation energy of the considered fragment. Following the definition introduced in Ref.~\cite{82Se_FRIB}, the $\Delta BE_{\mathrm{max}}$ and $S_n$ values are given in MeV and treated as numerical without explicit normalization within the logarithmic terms. The coefficients $k_1$, $k_2$, and $b$ are fitted parameters.

By taking into account a fragment's binding energy and one-neutron separation energy, these systematics become highly sensitive to nuclear structure effects near the neutron dripline. The mass systematics can be related to the experimentally measured cross sections by the relationship

\begin{equation}
    \sigma = 10^{\Delta BE_{\mathrm{sys}}}.
    \label{eq:dBEcs}
\end{equation}
where $\sigma$ is the production cross section in millibarns (mb). This relationship is used to determine the coefficients $k_1$, $k_2$, and $b$ in Equation~\ref{eq:dBE}.

Because $\Delta BE_{\mathrm{max}}$ and $S_n$ depend on the adopted mass model, the residuals about Equation~\ref{eq:dBEcs}, after fitting the systematics parameters, provide a measure of how consistently a given mass model describes the observed production cross-section trends. By extension, the systematics-predicted cross section value can be highly dependent on the chosen mass table.

\begin{figure}
    \centering
    \includegraphics[width=1.0\linewidth]{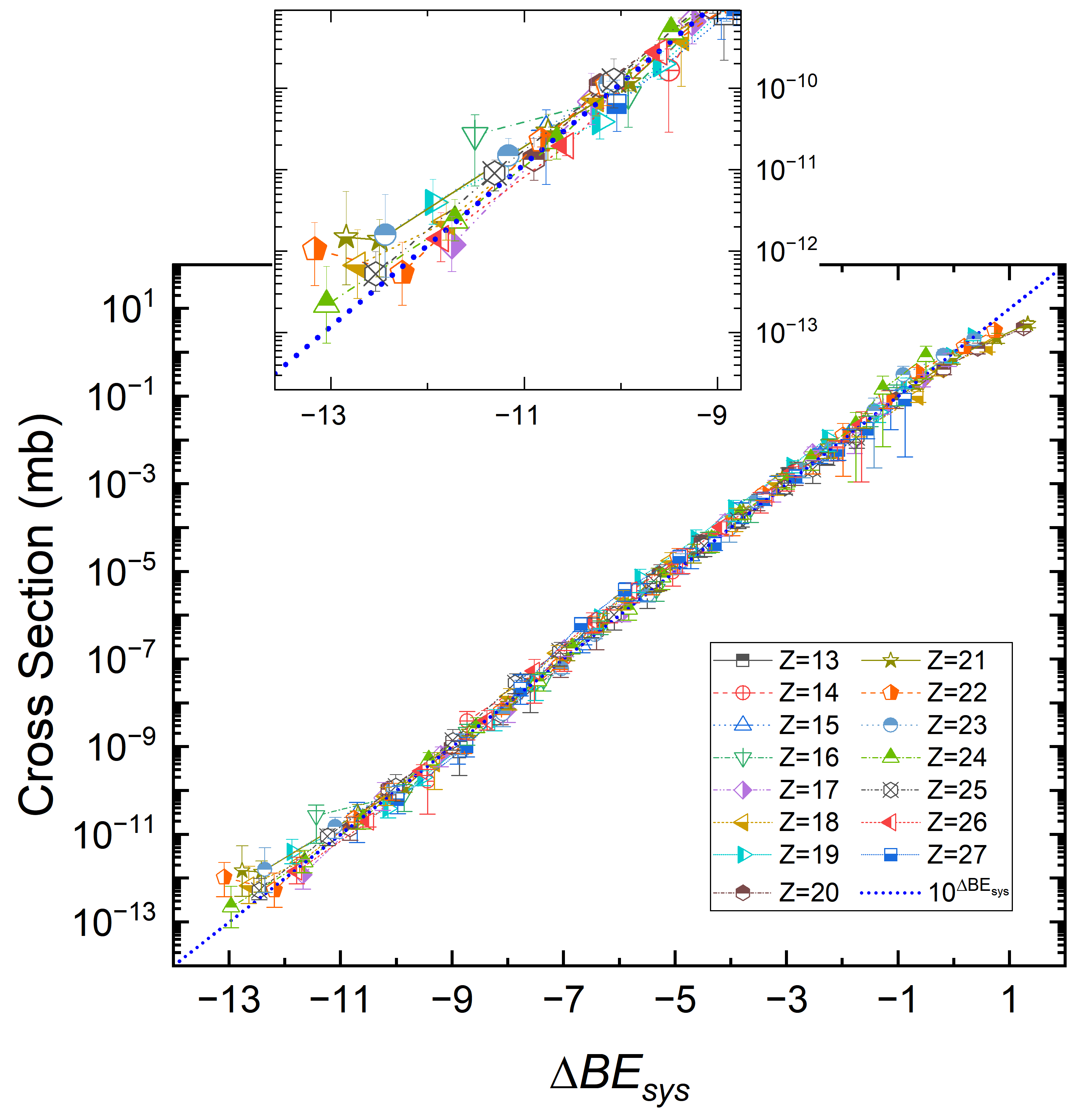}
    \caption{Cross sections from the fragmentation of \nuc{82}{Se} for neutron-rich isotopes of elements $13\leq Z \leq 27$ as a function of $\Delta BE$ systematics using the HFB-22 mass table. The fitted parameters for the systematics are provided in Table~\ref{tab:dBE}. The blue dotted line represents the function $\sigma= 10^{\Delta BE_{sys}}$. The inset shows the systematics of the most neutron-rich isotopes, and their deviation from the trendline. The deviation of the less neutron-rich isotopes is explained in the main text. Note that the parameters $k_1$ and $k_2$ are fixed within each $Z$ region, while the offset $b$ is fitted independently for each isotopic chain.}
    \label{fig:dBE}
\end{figure}

The result of fitting the $\Delta BE$ systematics to the measured cross sections, based on the HFB-22 mass table, is shown in Figure~\ref{fig:dBE}. The isotopic lines $Z=13-27$ are shown as various symbol and line types. The dotted blue line corresponds to $\sigma = 10^{\Delta BE_{\mathrm{sys}}}$ and is shown for reference. The determined coefficients of HFB-22 are summarized in Table~\ref{tab:dBE}. As with the AA calculations, HFB-22 provided a good description across the wide range of $Z$ considered. Note that since $b$ is an offset parameter, it was fit independently for each isotopic line. The $b$ values in Table~\ref{tab:dBE} represent a simple average over the $Z$ regions listed. The parameters determined for other mass models and the individual $b$ parameters used for each element are given in Appendix A in Tables~\ref{tab:dBE_allCoef} and \ref{tab:dBEoffset}, respectively.

A total of 183 cross sections of neutron-rich isotopes for elements $13\leq Z \leq 27$ were used in the minimization of the $\Delta BE$ systematics, grouped into three sets. The grouping was determined by the quality of the fit. Broadly, one set of parameters well describes the low-$Z$ isotopes, and a second set well describes the high-$Z$ isotopes. A small intermediate region could not be adequately described by either the low- or high-$Z$ parameterization and was therefore fitted separately. This intermediate region was observed in all the mass models considered. The boundary of the intermediate region is slightly model-dependent. SV-min and UNEDF0 have an intermediate region spanning $Z=18-21$; for the remaining  mass tables, the intermediate region  spans only $Z=18-19$.

As seen in Figure~\ref{fig:dBE}, the HFB-22 mass table provides a good description of the measured cross sections over a broad range of $Z$. Deviations from the expected trend for the most neutron-rich isotopes arise from disagreements between the measured values and the predicted cross sections based on the mass model. The largest deviations correspond to $^{63}$Sc and $^{66}$Ti. If the calculated binding-energy systematics for these nuclides are accurate, they suggest that the cross section should be much lower. For all eleven mass models considered, this observation was consistent.

The mass systematics provide an alternative explanation for the enhanced production cross section: that these nuclei have binding energies that differ significantly from those predicted by all models. All else held constant, the calculated systematics value can be pushed toward the trendline as the individual binding energy value is changed.

We stress here that modifying individual binding energies can lead to unphysical trends in the mass surface, specifically one-neutron separation energy. Investigating mass surface changes based on the $\Delta BE_{\mathrm{sys}}$ requires further study.

As mentioned previously, the measurement of $^{63}$Sc and $^{66}$Ti may have been contaminated by multistep reactions. These reactions, where the projectile undergoes sequential fragmentation in the primary target, can enhance the production rate of rare nuclei. The $\Delta BE$ systematics may provide information about multistep reactions in the production of neutron-rich nuclei \cite{82Se_FRIB}. A systematic study of this implication will be addressed in a subsequent paper.

The $\Delta BE$ systematics are only applicable for isotopes where the one-neutron separation energy influences the production cross section. For fragments closer to stability, this structural observable plays less of a role in the creation of the fragment. Therefore, deviations from the trend for the least neutron-rich isotopes are expected. This ``drooping'' of the line is observed in the top right of Figure~\ref{fig:dBE}.

\begin{table}[ht]
\centering
\caption{Coefficients and residual metric $R_{\Delta BE}$ of the $\Delta BE$ systematics for different mass regions, deduced from fits to the \nuc{82}{Se} cross section datasets using the HFB-22 mass table. The coefficients $k_1$, $k_2$, and $b$ correspond to those in Eq.~\ref{eq:dBE}. The offset parameter shown is a simple average over the $Z$ regions. Individual offsets are tabulated in Table~\ref{tab:dBEoffset}}

\setlength{\extrarowheight}{3pt}
\setlength{\tabcolsep}{7.0pt} 
\begin{ruledtabular}
\begin{tabular}{c|@{\hspace{-10pt}}c@{\hspace{6pt}}|c c c}
$Z$ Region & $R_{\mathrm{\Delta BE}}$ & $k_1$ & $k_2$ & $\left<b\right>$   \\ [3pt] \hline
$Z=13$--$17$ & 0.584 &  4.503 &  0.434 & -19.62 \\ 
$Z=18$--$19$  & 1.953 &  5.195 &  0.762 & -23.92 \\
$Z=20$--$27$  & 0.866 & 6.431 & 0.611 & -29.88 \\ 
\end{tabular}
\end{ruledtabular}
\label{tab:dBE}
\end{table}

The $\Delta BE$ systematics were calculated for the same eleven mass models listed in Table~\ref{tab:comparison_AA}. Across the models considered, trends exist in the coefficient $k_1$ and the offset parameters $b$. The coefficient $k_1$, corresponding to the natural logarithm of the difference in binding energy $\Delta BE_{\mathrm{max}}$, decreases as a function of removed protons $\Delta Z=Z_{\mathrm{beam}}-Z_{\mathrm{frag}}$. The coefficient $k_2$, corresponding to the natural logarithm of the one-neutron separation energy, follows a similar trend, except for a discontinuity in the intermediate region between low- and high-$Z$. The change in the fitted $k_2$ values may reflect evolution of the binding energy surface as the $Z=20$ shell closure is approached.

\subsection{Bayesian-Inspired Model Averaging}

The abrasion-ablation results depend in part on the mass model used. Differences in predicted binding energies affect the predicted production cross sections, even after AA parameters are refit, especially for weakly bound nuclei. As seen in Table~\ref{tab:comparison_AA}, mass models can vary in their goodness-of-fit across regions of the chart of nuclides. To reduce the systematic dependence on a mass model, a Bayesian-inspired model averaging (BMA) of the cross sections was performed, as outlined in Ref.~\cite{OlegBMA}.


For each region, the models were weighted according to their deduced parameter $1/\mathcal{J}_{\mathrm{AA}}^2$. Models that performed better in the region (smaller $\mathcal{J}_{\mathrm{AA}}$) were more heavily weighted. The central value is taken to be the weighted mean and the uncertainties are the 16$^{\text{th}}$ and 84$^{\text{th}}$ percentiles of the weighted distribution of model predictions. Where the mass models' predictions agree, the uncertainty range is small.

In order to perform the average, the cross sections were transformed into log space to prevent outliers (both large and small) from skewing the weighted results. After the average is taken, the cross section is returned to linear space.

Not every mass model agrees on which nuclei are bound. Models that predict a particular isotope to be unbound are excluded from the average for that isotope. The remaining weights are renormalized. The removal of the unbound model(s) inherently favors more optimistic models.

Individual mass tables rarely report error bars associated with individual binding energies. This makes the quantification of error bars associated with the $\Delta BE$ systematics for a single mass table difficult. To overcome this limitation, a Bayesian-inspired model averaging approach to the predicted cross sections of the $\Delta BE$ systematics was also employed.

The same eleven mass models used in the AA analysis were used for the $\Delta BE_{\mathrm{sys}}$ analysis. The BMA weighting is based on the  residual metric $R_{\mathrm{\Delta BE}}$, which includes a contribution from a penalty function if a mass model predicts that an observed isotope is unbound. The residual metric has the form

\begin{subequations}\label{eq:dBEresidual}
\begin{equation}
R_{\Delta BE} = \frac{1}{N-m}\sum_{i=1}^{N} r_i,
\end{equation}

\begin{equation}
r_i =
\begin{cases}
\left(\dfrac{D_i}{\delta_i}\right)^2,
    & S_{n,i}>0, \\[8pt]
100(S_{n,i}-\alpha)^2,
    & S_{n,i}\leq 0,
\end{cases}
\end{equation}

\begin{equation}
\delta_i =
\begin{cases}
\log_{10}\left(1+\dfrac{d\sigma_i^+}{\sigma_i}\right),
    & D_i\geq 0, \\[8pt]
-\log_{10}\left(1-\dfrac{d\sigma_i^-}{\sigma_i}\right),
    & D_i<0.
\end{cases}
\end{equation}
\end{subequations}
where $N$ is the number of fitted points, $m$ is the number of free parameters, $\sigma$ is the measured production cross section, $\Delta BE_{\mathrm{sys}}$ is the evaluated systematics value, $\alpha=0.01$ is a penalty parameter, and $D_i=\log_{10}(\sigma_i)-\Delta BE_{\mathrm{sys},i}$. The value of the residual metric depends on whether the calculated $\Delta BE_{\mathrm{sys},i}$ lies above or below the measured cross section $\sigma_i$. This is done to incorporate the asymmetric error bars of the production cross sections. In the penalty function, the size of the coefficient out front and the penalty parameter were set to ensure that the penalty scale was commensurate with the residual metric. When a mass model predicts that a particular isotope is unbound, the residual metric is undefined. The penalty function is then evaluated and replaces the undefined residual metric. The size of the penalty is based on the mass models' tabulated neutron separation energy. The larger the unbound prediction, the larger the penalty. The residual metric for $Z=17-27$ for all mass models is tabulated in Table~\ref{tab:comparison_AA}.

Models with a smaller residual metric across $Z=17-27$ were more heavily weighted in the BMA, according to $(1/R_{\Delta BE})^2$. Note that one weight was used across the full $Z$ range. The deduced model weights for both the AA and $\Delta BE_{\mathrm{sys}}$ calculations are shown in Table~\ref{tab:weights_AA_dBE}.

\begin{figure*}
    \centering
    \begin{minipage}{1.0\textwidth}
        \includegraphics[width=0.8\linewidth]{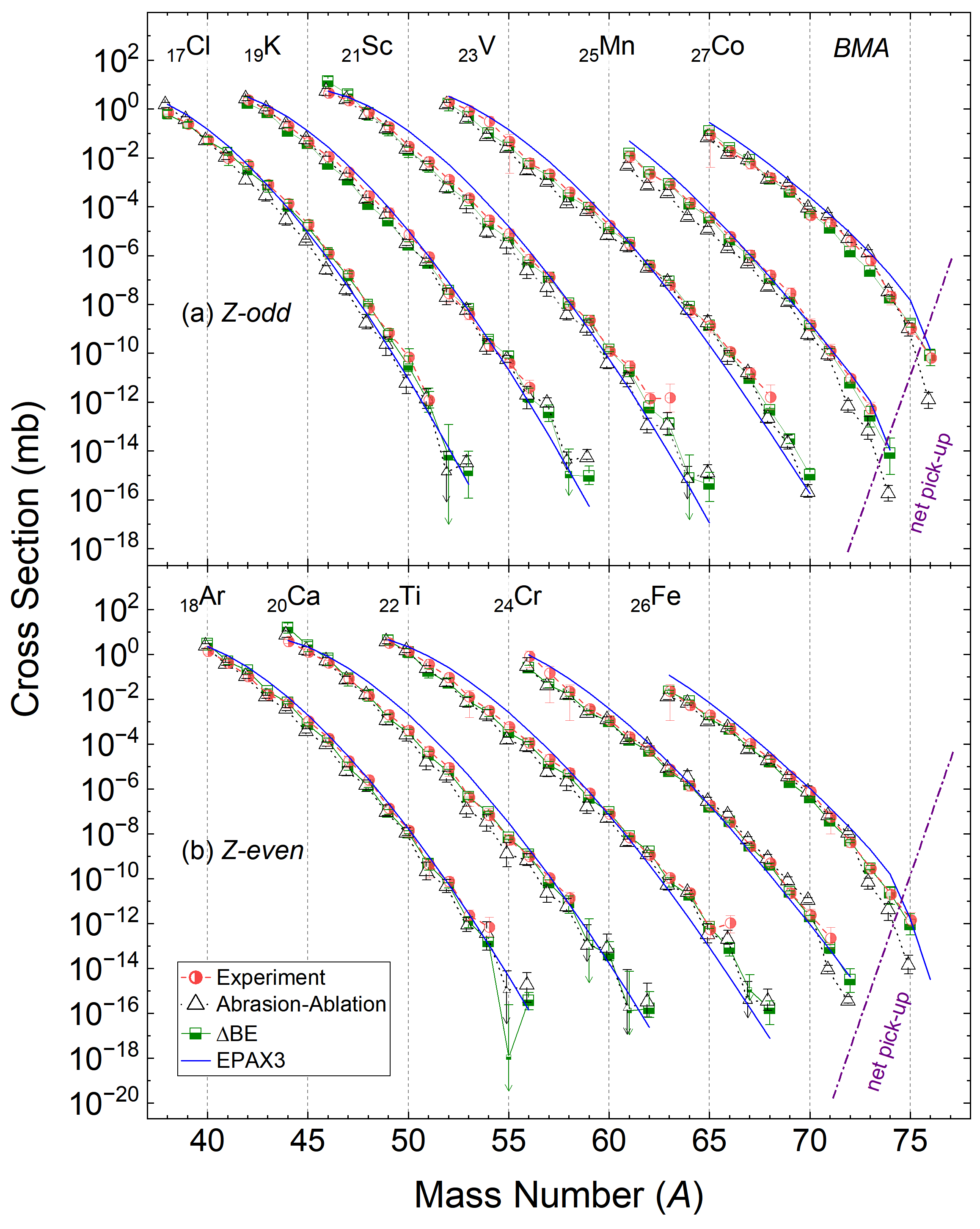}
        \caption{Comparison of experimentally measured cross sections from this work and from Refs.~\cite{MSU13,82Se_FRIB} (red half circles), abrasion-ablation calculations (open black triangles) based on a Bayesian-inspired model averaging of eleven mass models, predicted cross sections from the $\Delta BE$ systematics (green half squares) based on a Bayesian-inspired averaging of eleven mass models, and EPAX3~\cite{EPAX3} predictions (solid blue lines). Isotopes resulting from net neutron pick-up are denoted by the dark purple dot-dash line. The size of the symbol is proportional to the normalized ensemble weight associated with bound predictions. For any isotope with a model-weighted bound fraction less than one ($f_{\mathrm{bound}}<1$), a downward arrow is placed on the lower error bar.}
        \label{fig:CS_AA_dBE_EPAX3}
    \end{minipage}
\end{figure*}

When a nucleus has a negative one-neutron separation energy, the $\Delta BE$ systematics are undefined. A neutron-unbound nucleus corresponds to a zero measurable cross section for in-flight separators when the time-of-flight is significantly longer than the lifetime of the nucleus. The models which predict a particular nucleus to be unbound are excluded from the average for that nucleus. The remaining model weights are renormalized. This approach inherently favors more optimistic models. The central value and error bars were then converted into cross-section predictions through the relationship in Equation~\ref{eq:dBEcs}.

The BMA cross-section predictions from both the abrasion-ablation calculations and $\Delta BE$ systematics are shown in Figure~\ref{fig:CS_AA_dBE_EPAX3}. Note that the symbol sizes of the AA and $\Delta BE_{\mathrm{sys}}$-predicted cross sections are proportional to the model-weighted bound fraction $f_{\mathrm{bound}}$, which is the sum of the ensemble weights associated with models that predict the isotope to be bound. This model-weighted bound fraction is determined independently for AA and $\Delta BE_{\mathrm{sys}}$ from the weights of the underlying mass models. For isotopes with $f_{\mathrm{bound}}<1$, a downward-pointing arrow marks the lower uncertainty bound. This is to indicate that, while the error bars constitute the 16$^{\text{th}}$ and 84$^{\text{th}}$ percentiles of the remaining  cross-section predictions, the measured cross section may be significantly lower or zero.


\begin{table}[ht]
\setlength{\abovecaptionskip}{20pt} 
\centering
\setlength{\extrarowheight}{2pt}
\caption{BMA ensemble weights for abrasion-ablation calculations for \nuc{82}{Se} projectile-fragmentation products in the mass ranges $17 \leq Z \leq 19$ (Low), $20 \leq Z \leq 24$ (Medium), and $25 \leq Z \leq 27$ (High), and $\Delta BE$ systematics.}

\begin{ruledtabular}
\begin{tabular}{l@{\hspace{0.15cm}}|
                >{\centering\arraybackslash}p{0.85cm}
                >{\centering\arraybackslash}p{0.95cm}
                >{\centering\arraybackslash}p{0.85cm}
                |l}
\multirow{2}{*}{\raisebox{-12pt} {\parbox[c]{1.2cm}{\centering \hspace*{15pt}Mass\\\hspace*{15pt}Model}}} & 

\multicolumn{3}{c|}{ AA Weights (\%)} & \multirow{2}{*}{\raisebox{-12pt} {\parbox[c]{1.0cm}{\centering \hspace*{-0.45cm} $\Delta BE_{sys}$\\ \hspace*{-0.45cm} Weight \\ \hspace*{-0.45cm} (\%)}}} \\[-3pt]
 & \multicolumn{3}{c|} {\hspace*{-10pt}\rule{0.43\linewidth}{0.4pt}} \\
 & Low & Medium & High & \\[3pt]
\hline
AME2020         & 3.36  & 8.87 & 7.53  & 4.11    \\ 
FRDM2012            & 6.18   & 7.85 & 13.52 & 7.18     \\ 
HFB-22        & 18.94   & 22.85 & 11.61 & 30.76     \\ 
HFB-27            & 7.56   & 9.79 & 6.48 & 23.93    \\ 
KTUY         & 7.01   & 9.15 & 5.81 & 15.52     \\ 
LISE-LDM3       & 0.06   &  3.67 & 9.00 & 2.78    \\ 
SV-MIN        & 16.36   & 10.19 & 7.34 & 5.80   \\ 
TUYY     & 3.55   & 4.87 & 11.55 & 2.53     \\
UNEDF0           & 27.64   &  16.64 & 10.60 & 4.19    \\
UNEDF2           & 8.32   &  2.90 & 8.06 & 0.72   \\ 
WS4$_{RBF}$           & 1.02   & 3.24 & 8.50 & 2.48   \\ 
\end{tabular}
\end{ruledtabular}
\label{tab:weights_AA_dBE}
\end{table}




\section{Discussion}



\subsection{AA and \texorpdfstring{$\Delta BE_{\mathrm{sys}}$}{dBE} Model Comparison}

\begin{figure*}[t]
\centering
\includegraphics[width=1.0\textwidth]{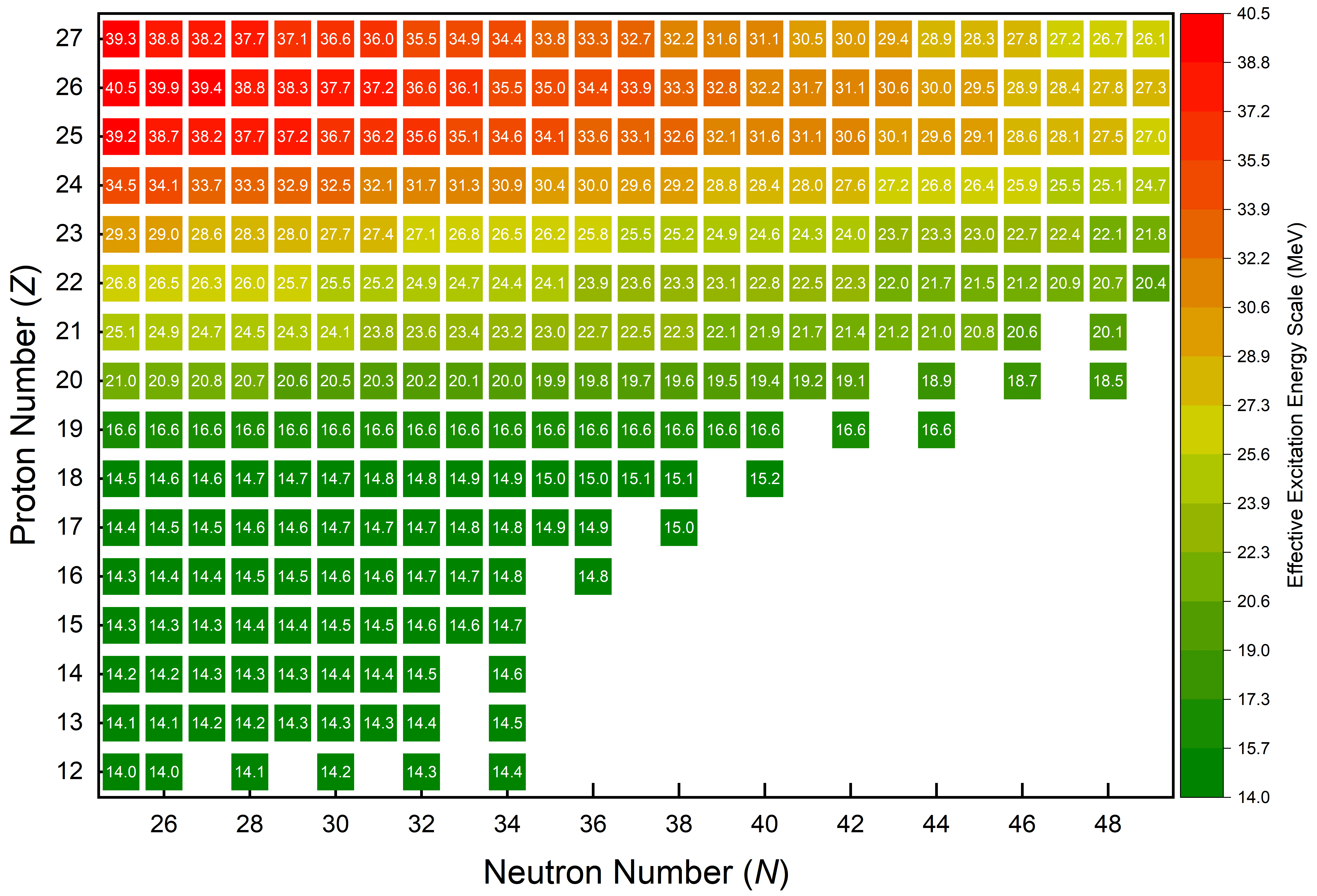}
\caption{\label{fig:BMAExcitationMap}
BMA-weighted effective excitation-energy scale parameter associated with the neutron-rich final fragments of $^{82}$Se with $12\leq Z\leq27$. The horizontal and vertical axes give neutron number $N$ and proton number $Z$, respectively. Numbers inside the isotope cells give the calculated value to one decimal place. To avoid artificial discontinuities associated with the three independently fitted $Z$ regions, the BMA coefficients were smoothly blended in $Z$ before evaluating the effective excitation-energy scale parameter, as described in Appendix B. The values for $Z=12$--16 are extrapolative because the abrasion-ablation parameters were constrained by experimental data only for $Z\geq17$. The values shown correspond to the production of the indicated final fragments and are not direct excitation energies of individual pre-fragments immediately after abrasion. For nuclei beyond the experimentally known region, the displayed isotope extent follows the UNEDF0 bound-nucleus predictions.}
\end{figure*}

There is good agreement between the cross sections predicted by AA and $\Delta BE$ systematics and the experimental data, demonstrating that the underlying components of the BMA weighted ensembles are consistent with the calibration data. It is notable that both the AA calculations and the $\Delta BE$ systematics extrapolations predict similar cross sections for unobserved nuclei. Deviations between the predictions arise primarily for nuclei that are predicted to be very weakly bound ($f_{\mathrm{bound}}<0.5$). For the heaviest elements studied (chromium-cobalt), large differences between the predictions are observed for nuclei near the pure-$Z$ removal line. The under-prediction of the abrasion-ablation model for these nuclides is an indication that other interactions with the target nuclei, such as charge exchange, may be missing from the current implementation of abrasion-ablation in \liseppsh.

The extrapolative agreement between the sets of calculations is an encouraging sign that the mass systematics, which are less computationally expensive to perform, can provide useful cross-section extrapolations for unobserved nuclei, subject to the systematic uncertainty associated with the choice of mass model.

More interesting than the discrepancies between the two approaches are cases in which they agree with one another but disagree with the measured cross section. This contradiction appears for the kinked nucleus $^{63}$Sc$_{42}$. Both AA and $\Delta BE$ systematics exhibit odd-even staggering in the vicinity of the kink, but fail to reproduce the higher measured cross section of the more neutron-rich isotope. 


Although the AA and $\Delta BE$ approaches differ in their implementation, both are sensitive to nuclear binding. The $\Delta BE$ systematics were developed from insights obtained with the AA model and provide an empirical description of production cross sections in terms of binding-energy differences and neutron separation energies. In AA calculations, nuclear masses influence particle-emission thresholds and competing decay widths during statistical de-excitation. While AA explicitly models pre-fragment formation, excitation energy, and particle evaporation, the $\Delta BE$ systematics offer a simpler parameterization of the resulting cross-section trends. The approaches are therefore complementary, but not physically independent.

The quantitative agreement between the two models with 
respect to the outlying data point indicates that the enhancement may arise from multistep reactions in the target or from deficiencies in the underlying mass model. Further examination to disentangle these effects is warranted and will be investigated in a subsequent paper.

The overall consistency of the calculation-intensive AA results
with the simpler $\Delta BE$ systematics results indicates that the systematics can provide useful cross section extrapolations toward the neutron dripline. As mentioned in Section~\ref{dBEsys}, the predicted cross section depends strongly on the adopted mass model; thus it becomes necessary to consider many models and the variation between them.

The measurement of \nuc{71}{Cr} provides an out-of-sample test of the predictive capability of the $\Delta BE$ systematics.  In Ref.~\cite{82Se_FRIB}, the $\Delta BE$ systematics were fitted to the available experimental data and used to predict production cross sections for previously unobserved isotopes.  Based on the $\Delta BE$ systematics coefficients reported in Tables III and IV of Ref.~\cite{82Se_FRIB} and the HFB-22 mass table, the predicted cross section of \nuc{71}{Cr} was $1.94\times10^{-13}$ mb. This is in good agreement with the measured value reported here: $2.20^{+2.10}_{-0.75}\times10^{-13}$ mb. The previously published and new cross-section predictions of  \nuc{69}{V} agree: $\approx 4\times 10^{-14}$ mb. Thus, \nuc{69}{V} is a promising candidate for the next isotope to be discovered in this region in an experiment of similar design.


\subsection{Effective Excitation-Energy Parameterization}

To describe the breadth of data analyzed in the abrasion-ablation framework accurately, the data were divided into three regions: low-$Z$ ($Z=17-19$), medium-$Z$ ($Z=20-24$), and high-$Z$ ($Z=25-27$). Each region has its own effective excitation energy coefficients, shown in Table~\ref{tab:AAcoef} in Appendix A. The relationship between the effective excitation energy scale parameter and the coefficients $c_1$, $c_2$, and $c_{NZ}$ is given in Equation~\ref{eq:eE_calc}. The change within and across these regions is of interest to ensure continuity and glean insight into the trends. As with the production cross sections, a BMA-inspired weighting of the parameters was performed for each model. The weights for each mass model are shown in Table~\ref{tab:weights_AA_dBE}; the resulting BMA parameters are tabulated at the bottom of Table~\ref{tab:AAcoef} in Appendix A. The calculation of the values presented in Figure~\ref{fig:BMAExcitationMap} is discussed in Appendix B.

Figure~\ref{fig:BMAExcitationMap} shows the effective excitation energy scale of neutron-rich fragments relevant to this experiment and future studies in the sub-calcium dripline region. The values shown, rounded to one decimal place, display a consistent trend: as the proton number of the final fragment decreases relative to the initial projectile, the effective excitation energy scale parameter decreases. This behavior reflects the variation of the fitted AA coefficients across the different regions.


Importantly, $\Delta A_{\mathrm{abr}}$, $\Delta Z_{\mathrm{abr}}$, and $\Delta N_{\mathrm{abr}}$ in Equation~\ref{eq:eE_calc} were replaced with the net differences in mass, proton, and neutron numbers between the projectile ($Z=34$, $N=48$) and each final fragment. These net differences do not necessarily correspond to the nucleons removed during the abrasion stage because the final fragments are populated through subsequent de-excitation and, in some cases, charge exchange.

As such, the values shown in Figure~\ref{fig:BMAExcitationMap} should be interpreted as a phenomenological map of the fitted excitation-energy scale parameterization, rather than as excitation energies of individual pre-fragments or averages over their production pathways.

A more direct interpretation would require an analysis of the excitation energy distributions and contributions of the individual abrasion pre-fragments that populate each observed isotope.



Understanding the trends in pre-fragment excitation energy is important for predicting the production of isotopes far from the primary beam. Recent measurements of \nuc{45,46}{Si} \cite{46Si} from a \nuc{70}{Zn} beam show the production cross sections are significantly higher than expected by EPAX2. Constraining the pre-fragment excitation energy across a broad range of fragments could provide valuable insight into the reaction mechanisms following abrasion.




\subsection{\texorpdfstring{$N-2Z$}{N-2Z} Investigation}

\begin{figure}
    \centering
    \includegraphics[width=1.0\linewidth]{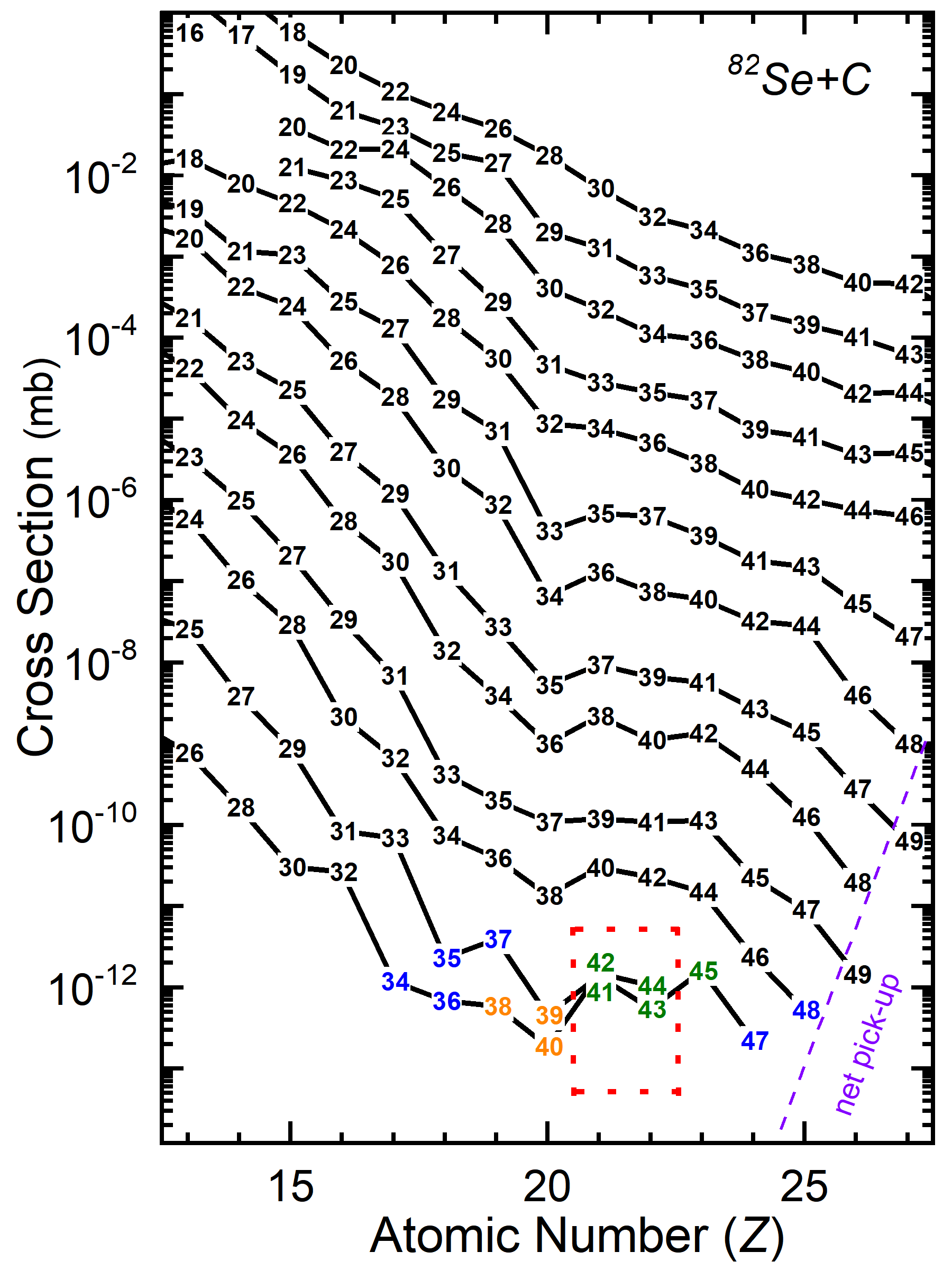}
    \caption{Production cross sections versus atomic number ($Z$) for fragments of \nuc{82}{Se}. Lines are connected along constant $N-2Z$, where labels are the neutron number. Figure adapted from Ref.~\cite{MSU13}. Note that labels for $^{62, 63}$Sc$_{41, 42}$ and $^{65, 66}$Ti$_{43, 44}$ have been shifted for visibility. Reactions resulting in net neutron pick-up are denoted by the dashed purple line. The colored values indicate cross sections measured in experiments subsequent to Ref.~\cite{MSU13}. The blue, green, and orange labels correspond to cross sections first measured in this work, Refs.~\cite{82Se_FRIB}, and \cite{60Ca} respectively. Error bars have been suppressed on this plot for clarity. The red dashed box highlights the crossing of the $N-2Z=0$ and $N-2Z=-1$ lines.}
    \label{fig:Nm2Z}
\end{figure}

Figure~\ref{fig:Nm2Z} displays the production cross sections from \nuc{82}{Se} along lines of constant $N-2Z$ as a function of $Z$, where the labels indicate $N$. This figure has been adapted from Ref.~\cite{MSU13}. The colored values indicate cross sections measured in experiments subsequent to Ref.~\cite{MSU13}. The blue, green, and orange labels correspond to cross sections first measured in this work, Refs.~\cite{82Se_FRIB}, and \cite{60Ca} respectively. The cross-section error bars have been suppressed for clarity. Cross sections resulting from net neutron pick-up are denoted by the dashed purple line.

For the most neutron-rich isotopes of scandium and titanium, the sharp kink in cross sections is manifested as a crossing of the $N-2Z=-1$ and $N-2Z=0$ lines. The crossing is highlighted by the red dashed box. Note that the labels for $^{62,63}$Sc$_{41, 42}$ and $^{65,66}$Ti$_{43, 44}$ have been shifted for clarity and that the ordering is based on their reported central values.

Beyond the $Z=20$ shell closure, the cross sections exhibit a pronounced odd-even staggering associated with neutron pairing: even-$N$ isotopes generally have enhanced production relative to their odd-$N$ neighbors. This staggering is weaker below the shell closure. The cross-section behavior of the most neutron-rich scandium and titanium isotopes, as well as the large gap that separates the colored labels from the black labels at and beyond $Z=20$, hint at potentially interesting nuclear structure effects in this region of the chart of nuclides. 


An additional feature of Figure~\ref{fig:Nm2Z} is the sudden drop in production cross section between \nuc{50}{Cl} and \nuc{51}{Cl}. 
Normally, large decreases in cross section occur when an unpaired neutron is added to the nucleus, especially after crossing a (sub)shell gap. In this case, adding a neutron produces an even-$N$ isotope whose cross section is strongly suppressed.  The suppression of \nuc{51}{Cl} makes this isotope an interesting candidate for nuclear-structure studies, particularly in connection with the possible evolution of the $N=34$ subshell closure below calcium \cite{expN34,LEBIT_N32_34,N34subShell}. Additionally, \nuc{51}{Cl} is predicted to be within a few neutrons of the dripline.


\section{Conclusion and Summary}

Production cross sections for neutron-rich nuclei produced in the fragmentation of \nuc{82}{Se} were measured for isotopes in the $17\leq Z \leq 27,\,\, 33\leq N \leq 49$ region, including the previously unobserved isotope \nuc{71}{Cr}. These measurements, combined with previous \nuc{82}{Se} fragmentation data, provide an extensive set of cross sections with which to investigate both the reaction mechanisms governing rare-isotope production and the influence of nuclear binding on the production of nuclei near the dripline.

The combined cross-section data also show an unusual suppression of \nuc{51}{Cl} relative to \nuc{50}{Cl}, contrary to the expected neutron odd-even staggering, motivating further investigation of nuclear binding near $N=34$.

Within the abrasion-ablation framework, systematic deviations between the measured and calculated cross sections point to limitations of the simple two-stage description of fragmentation. Multistep reactions in the production target, charge-exchange processes, and physics intermediate between abrasion and statistical ablation may contribute to these deviations. These observations indicate that additional reaction dynamics may need to be incorporated during and after the abrasion step to fully describe the production of neutron-rich nuclei from fragmentation.

The $\Delta BE$ systematics relate production cross sections to nuclear binding. Their strong sensitivity to the underlying mass model demonstrates the importance of nuclear-structure information in predicting rare-isotope production, while also providing a means of extrapolating cross sections beyond the experimentally accessible region. The first observation of \nuc{71}{Cr} provides an out-of-sample test of this approach: a cross section of $1.94\times 10^{-13}$ mb, predicted using the HFB-22 mass table and $\Delta BE_{\mathrm{sys}}$ coefficients determined before the observation of \nuc{71}{Cr}, is consistent with the measured value of $2.20^{+2.10}_{-0.75}\times10^{-13}$ mb.

Together, the abrasion-ablation and $\Delta BE_{\mathrm{sys}}$ approaches provide complementary descriptions of rare-isotope production. The systematic variation among the eleven nuclear mass models was incorporated into the cross section predictions through a Bayesian-inspired model averaging procedure. In particular, the simplicity of the $\Delta BE$ systematics makes them well suited for rapid extrapolation toward increasingly neutron-rich nuclei, while the comparison with abrasion-ablation calculations provides insight into the reaction mechanisms that may limit such predictions. The agreement for \nuc{71}{Cr} supports the predictive utility of the $\Delta BE$ systematics at the $10^{-13}$ mb scale, while the remaining discrepancies for \nuc{63}{Sc}, \nuc{66}{Ti}, and nuclei near the pure-$Z$ removal line identify priorities for improving reaction calculations.


\section*{Acknowledgments}

This material is based upon work supported by the U.S. Department of Energy, Office of Science, Office of Nuclear Physics and used resources of the Facility for Rare Isotope Beams (FRIB) Operations, which is a DOE Office of Science User Facility under Award Number DE-SC0023633. This work was supported by the U.S. National Science Foundation under Grants No. PHY-23-10078, No. PHY-2209429, No. PHY-2110365, No. PHY-2412343 (Mississippi State), and No. PHY-1848177 (CAREER; Mississippi State); the U.S. Department of Energy, Office of Science, Office of Nuclear Physics, under Contracts No. DE-AC02- 06CH11357 (ANL), No. DE-AC05-00OR22725 (ORNL), No. DE-AC52-07NA27344 (LLNL), and Grants No. DE-FG02-96ER40983 (UTK), No. DE-FG02-94ER40834 (Maryland), No. DE-SC0016988 (TTU), No. DE-FG02-97ER41041 (UNC), and No. DE-FG02-97ER41033 (TUNL); the Office of Defense Nuclear Nonproliferation Research and Development within the U.S. DOE - NNSA; the National Nuclear Security Administration under Award No. DE-NA0003180 and No. DE-NA0004249 and the Stewardship Science Academic Alliances program through DOE Awards No. DOE-DE-NA0003906, No. DOE-DE-NA0004068, and No. DOE-DE-NA0004074; the National Science Foundation Major Research Instrumentation Program Award No. 1919735 (UTK, TTU); the Laboratory Directed Research and Development Program at Pacific Northwest National Laboratory operated by Battelle for the U.S. Department of Energy; the Laboratory Directed Research and Development Program at Lawrence Livermore National Laboratory (Grant No. 23-LW-25); the Japan Society for the Promotion of Science, Grants-in-Aid for Scientific Research (KAKENHI) under Grant No. 22K14053.


\section*{Data Availability}

The experimental data and BMA cross section results for the abrasion-ablation and $\Delta BE_{\mathrm{sys}}$ are available at Ref.~\cite{Data}.

\appendix

\section{Abrasion-Ablation and \texorpdfstring{$\Delta BE_{\mathrm{sys}}$}{dBE} Coefficients}

Table~\ref{tab:AAcoef} lists the final minimized coefficients used for the abrasion-ablation calculations in \lisepp for the eleven mass models considered in this work. The bottom rows indicate the weighted parameter values from a BMA-inspired ensemble averaging. The weights used are tabulated in Table~\ref{tab:weights_AA_dBE}. Across all the regions and mass models, there are consistent trends in the parameters: the mean excitation energy coefficients $c_1$ and $c_2$ decrease as a function of proton loss; $c_{NZ}$ increases as a function of proton loss; the effective Coulomb barrier decreases as a function of proton loss; the odd-even staggering increases as a function of proton loss; the break-up limiting temperatures do not move much from their default values.

As noted in the main text, general trends exist in the $\Delta BE$ systematics coefficients $k_1$, $k_2$, and $b$. The trend in $k_2$ exhibits a sharp discontinuity associated with the intermediate region.


\begin{table*}[t]
\footnotesize
\caption{\label{tab:AAcoef}Abrasion-ablation coefficients deduced from a \lisepp minimization routine. The coefficients $c_1$, $c_2$, and $c_{NZ}$ correspond to those in Eq.~\ref{eq:eE_calc}. The remaining coefficients, from left to right, are: a global cross section scaling factor; the size of the effective Coulomb barrier; the phenomenological odd-even staggering; and the break-up temperature for $A=50$, 150, and 250, respectively. The BMA coefficients are weighted averages over the eleven mass models using the region-dependent AA weights of Table~\ref{tab:weights_AA_dBE}; uncertainties in parentheses are the corresponding weighted standard deviations. The regions correspond to low-$Z$ ($Z=17-19$), medium-$Z$ ($Z=20-24$), and high-$Z$ ($Z=25-27$), respectively.}
\setlength{\extrarowheight}{3pt}
\begin{ruledtabular}
\begin{tabular}{ccccccccccc}
Mass Model & Region & $c_1$ & $c_2$ & $c_{NZ}$ & A & dR & $\Delta OE$ & $T_{050}$ & $T_{150}$ & $T_{250}$\\
 [3pt] \hline 
  & Low & 15.91 & -0.05 & 0.01 & 1.05 & 5.45 & 10.86 & 8.00 & 5.90 & 4.70 \\
AME2020 & Med & 17.53 & 0.29 & -0.06 & 1.19 & 6.23 & 9.61 & 7.98 & 5.83 & 4.72 \\ 
 & High & 18.08 & 0.98 & -0.29 & 1.00 & 6.68 & 6.15 & 8.00 & 5.90 & 4.70 \\[3pt] \hline

  & Low & 16.93 & -0.14 & -0.01 & 1.02 & 5.50 & 10.20 & 8.00 & 5.90 & 4.70 \\
FRDM2012 & Med & 17.59 & 0.31 & 0.09 & 1.20 & 5.84 & 7.96 & 8.08 & 5.90 & 4.70 \\ 
 & High & 19.07 & 0.99 & -0.28 & 1.35 & 6.36 & 6.66 & 8.00 & 5.90 & 4.70 \\[3pt] \hline
 
 & Low & 17.06 & -0.05 & -0.01 & 1.00 & 5.32 & 10.87 & 7.99 & 5.90 & 4.70 \\
HFB-22 & Med & 17.50 & 0.29 & 0.04 & 0.90 & 6.22 & 9.61 & 8.03 & 5.86 & 4.69 \\ 
 & High & 18.66 & 1.05 & -0.32 & 0.92 & 6.73 & 6.07 & 8.05 & 5.93 & 4.69 \\[3pt] \hline

  & Low & 17.05 & -0.08 & 0.05 & 0.87 & 6.01 & 10.94 & 8.01 & 5.90 & 4.70 \\
HFB-27 & Med & 17.49 & 0.28 & 0.00 & 1.47 & 6.20 & 9.37 & 8.03 & 5.86 & 4.70 \\ 
 & High & 18.50 & 0.98 & -0.39 & 1.27 & 6.70 & 6.07 & 8.01 & 5.92 & 4.70 \\[3pt] \hline

  & Low & 16.01 & -0.12 & 0.08 & 0.78 & 5.64 & 9.61 & 8.05 & 5.89 & 4.71 \\
KTUY & Med & 17.61 & 0.21 & 0.00 & 1.01 & 6.27 & 8.97 & 8.01 & 5.87 & 4.70 \\ 
 & High & 18.90 & 0.74 & -0.27 & 0.83 & 6.48 & 6.45 & 8.00 & 5.90 & 4.70 \\[3pt] \hline

   & Low & 17.41 & -0.09 & 0.09 & 1.03 & 5.98 & 9.90 & 8.00 & 5.90 & 4.70 \\
LISE-LDM3 & Med & 17.58 & 0.27 & 0.09 & 0.99 & 6.26 & 9.59 & 8.01 & 5.91 & 4.70 \\ 
 & High & 18.00 & 1.05 & -0.32 & 1.20 & 6.54 & 6.00 & 8.00 & 5.90 & 4.70 \\[3pt] \hline
 
 & Low & 17.12 & -0.05 & -0.02 & 0.95 & 5.70 & 10.26 & 8.00 & 5.90 & 4.70 \\
SV-MIN & Med & 17.28 & 0.31 & 0.06 & 0.86 & 6.13 & 9.82 & 8.02 & 5.89 & 4.71 \\
 & High & 18.83 & 0.73 & -0.26 & 0.82 & 6.70 & 6.06 & 8.00 & 5.90 & 4.70 \\[3pt] \hline

 & Low & 17.06 & -0.05 & -0.02 & 0.94 & 6.08 & 10.53 & 8.00 & 5.90 & 4.70 \\
TUYY & Med & 17.42 & 0.39 & -0.11 & 1.39 & 6.23 & 9.58 & 7.98 & 5.90 & 4.70 \\
 & High & 19.19 & 1.02 & -0.23 & 0.87 & 6.60 & 6.20 & 8.01 & 5.89 & 4.71 \\[3pt] \hline

 & Low & 17.57 & -0.05 & 0.03 & 0.82 & 5.51 & 8.08 & 8.16 & 6.05 & 4.62 \\
UNEDF0 & Med & 17.97 & 0.11 & -0.09 & 1.00 & 5.69 & 7.05 & 8.09 & 5.85 & 4.79 \\
 & High & 18.05 & 0.49 & -0.31 & 0.80 & 6.50 & 6.00 & 7.99 & 5.91 & 4.69 \\[3pt] \hline

 & Low & 16.11 & -0.11 & 0.12 & 1.25 & 6.07 & 10.00 & 8.03 & 5.84 & 4.70 \\
UNEDF2 & Med & 17.74 & 0.30 & 0.11 & 1.03 & 6.31 & 6.46 & 8.01 & 5.90 & 4.70 \\
 & High & 19.18 & 0.42 & -0.30 & 0.82 & 7.00 & 6.04 & 8.01 & 5.85 & 4.74 \\[3pt] \hline

 & Low & 16.01 & -0.05 & -0.01 & 1.31 & 5.32 & 10.86 & 8.03 & 5.96 & 4.70 \\
WS4$_{RBF}$ & Med & 17.54 & 0.31 & 0.04 & 1.30 & 6.40 & 7.97 & 8.00 & 5.92 & 4.71 \\
 & High & 18.71 & 0.70 & -0.31 & 0.91 & 6.47 & 6.00 & 8.00 & 5.90 & 4.70 \\[3pt] \hline

 & Low & 17.00(58) & -0.068(34) & 0.022(46) & 0.94(14) & 5.61(27) & 9.79(1.22) & 8.05(7) & 5.93(8) & 4.68(4) \\
BMA & Med & 17.58(21) & 0.262(79) & 0.005(70) & 1.07(21) & 6.11(23) & 8.85(1.15) & 8.03(4) & 5.87(2) & 4.72(4) \\
 & High & 18.67(46) & 0.847(233) & -0.296(38) & 0.99(21) & 6.60(18) & 6.17(23) & 8.01(2) & 5.90(2) & 4.70(1) \\
\end{tabular}
\end{ruledtabular}
\end{table*}

\begin{table}[ht]
\footnotesize
\centering
\caption{Coefficients and residual metric $R_{\mathrm{\Delta BE}}$ of the $\Delta BE$ systematics for different mass regions, deduced from fits to the \nuc{82}{Se} cross section datasets. The BMA coefficients are weighted averages over the eleven mass models using the $\Delta BE_{\mathrm{sys}}$ weights listed in Table~\ref{tab:weights_AA_dBE}; uncertainties in parentheses are the corresponding weighted standard deviations.}

\setlength{\extrarowheight}{3pt}
\setlength{\tabcolsep}{7.0pt} 
\begin{ruledtabular}
\begin{tabular}{c|@{\hspace{-4pt}}c@{\hspace{6pt}}|@{\hspace{-4pt}}c@{\hspace{5pt}}|c c }
Mass Model & $Z$ Region & $R_{\mathrm{\Delta BE}}$ & $k_1$ & $k_2$   \\ [3pt] \hline
& $13$--$17$ & 3.409 &  3.544 &  0.597 \\ 
AME2020 & $18$--$19$  & 1.857 &  3.457 &  0.971  \\
& $20$--$27$  & 2.198 & 5.176 & 0.901  \\ [3pt] \hline 

& $13$--$17$ & 1.720 &  3.608 &  0.643 \\ 
FRDM2012 & $18$--$19$  & 2.532 &  4.875 &  0.824  \\
& $20$--$27$  & 1.899 & 5.747 & 0.677  \\ [3pt] \hline 

& $13$--$17$ & 0.584 &  4.503 &  0.434 \\ 
HFB-22 & $18$--$19$  & 1.953 &  5.195 &  0.762  \\
& $20$--$27$  & 0.866 & 6.431 & 0.611  \\ [3pt] \hline 

& $13$--$17$ & 0.632 &  3.719 &  0.504 \\ 
HFB-27 & $18$--$19$  & 1.998 &  4.324 &  0.639  \\
& $20$--$27$  & 1.041 & 5.523 & 0.589  \\ [3pt] \hline 

& $13$--$17$ & 1.144 &  3.558 &  0.601 \\ 
KTUY & $18$--$19$  & 3.803 &  4.427 &  0.825  \\
& $20$--$27$  & 0.849 & 5.900 & 0.675  \\ [3pt] \hline 

& $13$--$17$ & 2.158 &  3.386 &  0.496 \\ 
LISE-LDM3 & $18$--$19$  & 4.634 &  3.873 &  0.255  \\
& $20$--$27$  & 3.178 & 5.037 & 0.746  \\ [3pt] \hline 

& $13$--$17$ & 2.721 &  4.295 &  0.573 \\ 
SV-MIN & $18$--$21$  & 2.771 &  4.938 &  0.632  \\
& $22$--$27$  & 1.412 & 6.249 & 0.592  \\ [3pt] \hline 

& $13$--$17$ & 2.993 &  3.341 &  0.662 \\ 
TUYY & $18$--$19$  & 7.736 &  5.057 &  0.607  \\
& $20$--$27$  & 2.320 & 6.163 & 0.708  \\ [3pt] \hline 

& $13$--$17$ & 2.732 &  4.666 &  0.827 \\ 
UNEDF0 & $18$--$21$  & 2.836 &  4.889 &  0.760  \\
& $22$--$27$  & 2.330 & 6.658 & 0.666  \\ [3pt] \hline 

& $13$--$17$ & 10.685 &  3.564 &  0.490 \\ 
UNEDF2 & $18$--$19$  & 2.980 &  3.710 &  0.578  \\
& $20$--$27$  & 4.641 & 6.210 & 0.660  \\ [3pt] \hline 

& $13$--$17$ & 6.095 &  3.210 &  0.540 \\ 
WS4$_{RBF}$ & $18$--$19$  & 2.465 &  2.984 &  0.842  \\
& $20$--$27$  & 2.166 & 4.724 & 0.683  \\ [3pt] \hline

& Low & -- & 3.96(51)  & 0.53(11)  \\ 
BMA & Interm.  & -- & 4.63(60)  & 0.73(14)   \\
& High  & -- & 5.94(53) & 0.64(8)  \\
\end{tabular}
\end{ruledtabular}
\label{tab:dBE_allCoef}
\end{table}

\begin{table*}[ht]
\footnotesize
\caption{\label{tab:dBEoffset}$\Delta BE$ systematics offsets $b$ deduced from a \lisepp minimization routine.}
\setlength{\extrarowheight}{3pt}
\begin{ruledtabular}
\begin{tabular}{cccccccccccc}
& \multicolumn{10}{c}{Mass Model}  \\ 
$Z$ & AME2020 & FRDM2012 & HFB-22 & HFB-27 & KTUY & LISE-LDM3 & SV-MIN & TUYY & UNEDF0 & UNEDF2 & WS4$_{RBF}$  \\
 [3pt] \hline 

  13 & -15.59 & -15.74 & -19.00 & -15.93 & -15.48 & -14.73 & -18.29 & -15.15 & -20.07 & -15.66 & -14.31 \\

  14 & -16.22 & -15.87 & -19.47 & -16.41 & -16.12 & -15.30 & -19.09 & -15.34 & -20.81 & -16.25 & -15.02 \\

  15 & -16.13 & -16.15 & -19.68 & -16.61 & -16.07 & -15.38 & -19.29 & -15.12 & -21.06 &  -16.46 & -15.00 \\

  16 & -16.25 & -16.54 & -19.88 & -16.77 & -16.06 & -15.43 & -19.40 & -15.50 & -21.13 & -16.44 & -14.91 \\

  17 & -16.00 & -16.85 & -20.06 & -16.68 & -16.32 & -15.33 & -19.23 & -15.91 & -21.09 & -15.81 & -14.48 \\

  18 & -16.76 & -21.88 & -23.61 & -19.54 & -20.03 & -16.89 & -22.11 & -22.07 & -22.30 & -16.97 & -14.60 \\

  19 & -17.19 & -22.26 & -24.24 & -19.97 & -20.74 & -17.43 & -22.40 & -23.14 & -22.45 & -17.21 & -14.98 \\

  20 & -23.46 & -26.16 & -29.21 & -24.62 & -26.37 & -22.36 & -23.09 & -28.21 & -23.17 & -26.02 & -20.66 \\

  21 & -24.04 & -26.62 & -29.55 & -25.02 & -26.82 & -23.15 & -23.49 & -28.58 & -23.60 & -27.27 & -21.38 \\

  22 & -24.93 & -26.92 & -29.76 & -25.39 & -27.22 & -23.95 & -28.41 & -28.86 & -30.04 & -28.25 & -22.07 \\

  23 & -24.97 & -26.78 & -29.95 & -25.71 & -27.23 & -24.08 & -28.82 & -28.64 & -30.60 & -28.74 & -22.29 \\

  24 & -24.97 & -26.88 & -30.14 & -25.85 & -27.54 & -24.10 & -29.30 & -28.66 & -31.17 & -29.02 & -22.53 \\

  25 & -24.63 & -26.67 & -30.07 & -25.83 & -27.59 & -23.88 & -29.37 & -28.44 & -31.37 & -29.18 & -22.54 \\

  26 & -24.49 & -26.57 & -30.27 & -25.96 & -27.77 & -23.80 & -29.48 & -28.46 & -31.50 & -29.15 & -22.64 \\

  27 & -24.64 & -26.37 & -30.12 & -25.80 & -27.48 & -23.65 & -29.17 & -28.29 & -31.28 & -28.82 & -22.51 \\

\end{tabular}
\end{ruledtabular}
\end{table*}

\section{Effective Excitation Energy Parameter Calculation}

The abrasion-ablation calculations were performed in three regions, chosen to provide an accurate description of the experimental data. These regions are low-$Z$ ($Z=17-19$), medium-$Z$ ($Z=20-24$), and high-$Z$ ($Z=25-27$). The coefficients of Equation~\ref{eq:eE_calc} were determined during the AA minimization. The BMA-weighted coefficients used to calculate $T_{\mathrm{eff}}$ are shown at the bottom of Table~\ref{tab:AAcoef} in Appendix A. The results are plotted in Figure~\ref{fig:BMAExcitationMap}. This appendix describes the procedure by which those values were calculated.

The parameter sets for each region were determined individually. This causes sharp discontinuities across the region boundaries. To avoid abrupt changes at these boundaries, the BMA coefficients were blended as a function of $Z$. The resulting values of $T_{\mathrm{eff}}$ were then calculated without further smoothing. Between $Z=18$ and 22, the Low and Medium coefficients were joined with the smoothstep weight $w(x)=3x^2-2x^3$, with $x=(Z-18)/4$. The same interpolation was applied between the Medium and High coefficients over $22\leq Z\leq26$, with $x=(Z-22)/4$. The interpolation is introduced only to visualize the systematic evolution of the fitted parameters and does not constitute an additional fit to the data.

For $Z<17$, the low region coefficients $c_1$, $c_2$ and $c_{NZ}$ were uniformly propagated. The abrasion-ablation parameters were constrained by experimental data only for $Z \geq 17$. Therefore, the extension of the BMA-weighted excitation-energy systematics to $Z=12$--16 is presented only as an approximate guide for the effective excitation energy scale for the light-$Z$ isotopes.

The resulting parameter values were then evaluated isotope by isotope according to Equation~\ref{eq:eE_calc} for the production of the final fragments; no additional smoothing was applied to the calculated values. $\Delta A_{\mathrm{abr}}$, $\Delta Z_{\mathrm{abr}}$, and $\Delta N_{\mathrm{abr}}$ in Equation~\ref{eq:eE_calc} were replaced with the net differences in proton, neutron, and mass numbers between the projectile ($Z=34$, $N=48$) and each final fragment. We reiterate that these net differences do not generally correspond to the nucleons removed during the abrasion stage because the final fragments are populated through subsequent de-excitation and, in some cases, charge exchange.

As mentioned in the main text, the calculated values should be interpreted simply as a phenomenological map of the fitted excitation-energy parameterization. The calculated values are not the excitation energies of individual pre-fragments.



%

\end{document}